\documentclass[aps,prx,twocolumn,superscriptaddress]{revtex4-1}

\usepackage[compat=1.0.0]{tikz-feynman}
\usepackage{amssymb}
\usepackage{multirow}
\usepackage{bm}
\usepackage{changes}
\usepackage{amsmath}
\usepackage{graphicx}
\usepackage{epstopdf}
\usepackage{subfigure}
\usepackage{natbib}
\usepackage{epsfig}
\usepackage{amsfonts}
\usepackage{mathrsfs}
\usepackage{braket}
\usepackage[toc,page,title,titletoc,header]{appendix}
\usepackage[colorlinks,linkcolor=blue,citecolor=blue,anchorcolor=blue]{hyperref}
\usepackage{dsfont,amsthm,amsbsy}

\usepackage{fancyhdr}

\usetikzlibrary{quotes,angles}
\newcommand{\obtuseangle}{\kern.08em
\begin{tikzpicture}
    \draw coordinate (a) at (0.14,0);
    \draw coordinate (b) at (0,0);
    \draw coordinate (c) at (-.12,0.18);
    \draw (a) -- (b) -- (c) pic [draw=black]{} ;
\end{tikzpicture}%
\kern.08em%
}

\newcommand{\be}{\begin{eqnarray}}
\newcommand{\ee}{\end{eqnarray}}

\begin{document}

\title{Superconducting valence bond fluid in lightly doped 8-leg $t$-$J$ cylinders}
\author{Hong-Chen Jiang}
\email{hcjiang@stanford.edu}
\affiliation{Stanford Institute for Materials and Energy Sciences, SLAC National Accelerator Laboratory and Stanford University, Menlo Park, California 94025, USA}

\author{Steven A. Kivelson}
\email{kivelson@stanford.edu}
 \affiliation{Department of Physics, Stanford University, Stanford, California 94305, USA}
 
\author{Dung-Hai Lee}
\email{dunghai@berkeley.edu}
\affiliation{Department of Physics, University of California, Berkeley, CA 94720, USA.\\
 Materials Sciences Division, Lawrence Berkeley National Laboratory, Berkeley, CA 94720, USA.
}
\date{\today}
\begin{abstract}

Superconductivity in doped quantum paramagnets has been a subject of long theoretical inquiry. In this work we report a density matrix renormalization group study of lightly doped $t$-$J$ models on the square lattice (doped hole densities $\delta = 1/12$ and 1/8) with parameters for which previous studies have suggested that the undoped system in  2D  is either a quantum spin liquid or a valence bond crystal. Our studies are performed on cylinders with width up to 8. Ground-state correlations are found to be nearly identical for the ``doped quantum spin liquid'' and ``doped valence bond crystal''. Upon increasing the cylinder width from 4 to 8, we observed a significant strengthening of the quasi-long-range superconducting correlations, and a dramatic suppression of any ``competing'' charge-density-wave order. Extrapolating from the observed behavior of the width 8 cylinders, we speculate that  the system has a nodeless d-wave superconducting ground-state in the 2D limit.
\end{abstract}

\maketitle

\section{Introduction}%
The mechanism by which superconductivity (SC) arises from doped ``Mott insulators''
continues to attract broad theoretical interest, especially as it relates to the mechanism of high temperature superconductivity in the cuprates\cite{Anderson1987,Lee2006,Weng2011,broholmreview}. Based on  extrapolations from weak coupling\cite{Arovas2022} or various mean-field theories\cite{Weng1999,Qin2022}, suggestive evidence has accrued that unconventional superconductivity emerges near half-filling in the Hubbard and related models with strong short-range electron-electron repulsion. However, controlled numerical treatments of the intermediate coupling problem, especially using density-matrix renormalization group (DMRG)\cite{AbsenceSC,White1999,scalapinowhitereview,Qin2022,Dodaro2017,Arovas2022}, 
have found that SC is less ubiquitous than was originally conjectured.\footnote{Similar conclusions concerning competing orders quenching SC were reached on the basis of variational auxiliary field quantum Monte Carlo calculations in Ref. \cite{sorella,ZhangStripes,Xiao2023}.} For example, in the ``pure'' Hubbard or the related $t$-$J$ model (with only nearest-neighbor (NN) hopping $t_1$), the undoped system (i.e. with $n=1$ electron per site) is well known to exhibit strong N\'eel antiferromagnetic (AF) order characterized by a NN exchange coupling $J_1 \approx 4t_1^2/U$. However, there is a growing consensus that unidirectional charge-density-wave (CDW) (i.e. ``stripe'') order rather than SC arises for doped hole concentrations, $\delta\equiv 1-n$, in the interesting range $0 < \delta < 1/4$.\cite{White1999,scalapinowhitereview,Dodaro2017,Zheng2017,Jiang2020prr,Gong2021,Jiang2021White,Jiang2018tJ,Jiang2020tJ,Chung2020} In the presence of next-nearest-neighbor (NNN) hopping $t_2$ and the generated exchange coupling $J_2$, the AF order at $\delta=0$ tends to be frustrated. At the same time, for $\delta >0$, SC correlations are found to be significantly enhanced,\cite{Jiang2019Hub,Jiang2020prr,Chung2020,Gong2021,Jiang2021White,Jiang2021,Peng2022} 
although, on systems wider than four-legs, this enhancement is only observed when $t_2 >0$ \cite{Gong2021,Jiang2021White,Jiang2021}. The dependence of SC on the sign of $t_2$ was surprising in the cuprate context, given that to reproduce the band dispersions of hole-doped cuprates seen in ARPES, i.e. to obtain a closed hole-like Fermi surface enclosing the $(\pi,\pi)$ point, requires $t_2$ negative.\cite{damascelli}

In this paper we study the $t$-$J$ model with doping concentrations $\delta=1/12$ and $1/8$ and with  parameters $t_1/J_1=3$ ($U_{\rm eff}\equiv 4t_1^2/J_1=12t_1$) and $t_2/t_1=\sqrt{0.5}$ and $\sqrt{0.55}$ such that $J_2/J_1=0.5$ and $0.55$. Like other DMRG studies, this is done on cylinders with a finite width $W$ and length $L > W$. The maximum $W$ we can study is $W=8$. (For $W=8$ we have considered $L=24$.) For these values of $J_2/J_1$, {earlier DMRG studies suggest that the undoped ($\delta=0$) system at $J_2/J_1=0.5$ is a quantum spin liquid (QSL), while at $J_2/J_1=0.55$ it is a valence bond crystal (VBC) in the 2D limit.\cite{Gong2014,Wang2018,Liu2022SciBul,Liu2022PRX} ({Other scenarios has also been proposed.}) However on cylinders with $W\le 8$ we found no qualitative difference between these two values of $J_2/J_1$. In both cases, the state can be characterized as a quantum paramagnet in the sense that  there is a finite spin-gap and  spin-spin (see section \ref{Sec:SG}) and dimer-dimer (see section \ref{Sec:Dimer}) correlations both fall exponentially with distance. Moreover, the correlation lengths are smaller than $W$ when $W\geq 6$ and show no tendency to increase with increasing $W$.}\cite{Jiang2012,Gong2014,Wang2018,Liu2022SciBul,Liu2022PRX}.

For $\delta=1/12$ and $1/8$, we find the ground state is a Luther-Emery liquid with superconducting quasi-long-range (power law decaying) order. Moreover, even though the cylinders break the crystal 90-degree rotation symmetry, the SC correlations we find are surprisingly isotropic which is similar to findings of previous studies.\cite{Jiang2021,Gong2021} The symmetry of the SC order parameter is d-wave. For large $W$, the exponent $K_{\rm sc}$ characterizing the power law decay of correlations is expected\cite{gannot} to decrease as $K_{\rm sc}\propto 1/W$;  while we find roughly comparable values of this exponent for $W=4$ and 6, it is roughly a factor of 2 smaller for $W=8$.  Moreover, the relatively short values of $\xi_{\rm AF}< W$ on the finite $(W,L)$ cylinders we have studied are suggestive that the existence of a spin-gap persists in the 2D limit.

However, despite the uncertainties concerning the half-filled ground state at these values of $J_2/J_1$, we believe our study provides plausible evidence that upon relatively light doping, the ground state becomes a nodeless d-wave superconductor. In addition to the already mentioned  isotropy of the SC correlations, this conclusion is supported by the lack of evidence of other orders in close competition with SC. In other words, the SC is locally stable, and has the right properties for being a finite cylinder manifestation of a 2D SC. Our belief is also based on the intuition that so long as the system is quantum paramagnetic at half filling, even if it has weak symmetry-breaking such as a VBC, when $\delta$ exceeds a (possibly small) critical value, and when the  hole hopping is sufficiently strong,  the moving holes will quantum melt the crystalline order of the dimers so that the resulting state is superconducting. (This is the same intuition as that of the short-range RVB picture\cite{rokhsarandme}. This intuition  is further supported by an earlier DMRG  study of a striped Hubbard cylinder \cite{Jiang2022pnas}), and a sign-problem free quantum Monte-Carlo simulation of lightly doped VBC on honeycomb lattice\cite{Li2023}.

The most salient results of our study can be summarized as follows: (1) We have extended an earlier DMRG study\cite{Jiang2021} with $J_2/J_1=0.5$, to include the value $J_2/J_1=0.55$, which, according to Ref.\cite{Gong2014,Wang2018,Liu2022SciBul,Liu2022PRX}, is in the VBC phase at half-filling. We increased the maximum width of the cylinders from $W=6$\cite{Jiang2021}, to $W=8$, which was computationally costly, requiring us to keep an enormous number -- up to $m=60,000$ - of states. For both values of $J_2/J_1$, we observed a quantum paramagnetic state with no apparent symmetry breaking at half filling. (2) When doped with $\delta=1/12$ and $1/8$ holes, the ground state exhibits d-wave SC quasi-long-range-order. The correlation function exhibits a high degree of isotropy, as expected for a 2D superconductor.  (3) For $W=8$ there is no other apparent order in close competition with superconductivity. (4) On the widest cylinders we studied ($W=8$), the estimated decay exponent ($K_{\rm sc}$) of the quasi-long-range-ordered superconductivity is around $0.5$. Within (significant) error bars this is nearly a factor of two smaller than the value ($\approx 1$) obtained for $W=4$ and 6 cylinders. ({We summarize various decay exponents in Table \ref{Table:Sum} and comment on the error bars.})

The paper is organized such that in the Results section we focus on presenting the numerical facts. The implications of these results are left to the Summary and Discussions section.

\section{Model and Method}\label{Sec:Method}
We employ DMRG\cite{White1992} to study the ground state properties of the hole-doped $t$-$J$ model on the square lattice, with Hamiltonian
\begin{eqnarray}
H=-\sum_{ij\sigma} t_{ij} \left(\hat{c}^\dagger_{i\sigma} \hat{c}_{j\sigma} + h.c.\right) + \sum_{ij}J_{ij}\left (\vec{S}_i\cdot \vec{S}_j - \frac{\hat{n}_i \hat{n}_j}{4} \right )\nonumber
\end{eqnarray}
where $\hat{c}^\dagger_{i\sigma}$ ($\hat{c}_{i\sigma}$) is the electron creation (annihilation) operator on site $i=(x_i,y_i)$ with spin polarization $\sigma$, $\vec{S}_i$ is the spin operator and $\hat{n}_i=\sum_{\sigma}\hat{c}^\dagger_{i\sigma}\hat{c}_{i\sigma}$ is the electron number operator. The electron hopping amplitude $t_{ij}$ is equal to $t_1$ ($t_2$) if $i$ and $j$ are NN (NNN) sites. $J_1$ and $J_2$ are the spin superexchange interactions between NN and NNN sites, respectively. The Hilbert space is constrained by the no-double occupancy condition, $n_i=0$ or 1. At half-filling, i.e., $n_i=1$, $H$ reduces to the spin-1/2 AF $J_1$-$J_2$ Heisenberg model.

We take the lattice geometry to be cylindrical with periodic and open boundary conditions in the $\hat{y}$ and $\hat{x}$ directions, respectively, where $\hat{y}=(0,1)$ and $\hat{x}=(1,0)$ are the two basis vectors of the square lattice. Here, we focus on cylinders with width $W$ and length $L$, where $L$ and $W$ are the number of sites along the $\hat{x}$ and $\hat{y}$ directions, respectively. The total number of sites is $N=L\times W$, the number of electrons $N_e$, and the doping level of the system is defined as $\delta=N_h/N$, where $N_h=N-N_e$ is the number of doped holes relative to the half-filled insulator with $N_e=N$. We set $J_1$=1 as an energy unit, and consider $J_2=0.5$ and $J_2=0.55$ such that according to Refs. \cite{Gong2014,Wang2018,Liu2022SciBul,Liu2022PRX} the undoped system in the 2D limit is, respectively, in the QSL and VBC  phases at half-filling. We take $t_1=3$, which to the extent that the results can be related to a corresponding Hubbard model, would correspond to $U^{\rm eff}\equiv 4t/J = 12t$.With relation to the Hubbard model in mind, we also impose the  condition
$t_2/t_1\ =\ \sqrt{J_2/J_1}$.

An advantage of DMRG is that large values of $L$ are computationally accessible, so we consider cylinders with $L \gg W$. We consider $W=4 - 8 $ cylinders at both $\delta = 1/12$ and $1/8$. We keep up to $m=16000$ states for $W=4$ cylinders with a typical truncation error $\epsilon< 10^{-8}$, $m=25000$ states for $W=6$ cylinders with a typical truncation error $\epsilon< 10^{-6}$, and $m=60000$ states for $W=8$ cylinders with a typical truncation error $\epsilon<10^{-5}$. Further details of the numerical simulation are provided in Appendices \ref{SM:SC} and \ref{SM:CDW} . In Figs.\ref{Fig:SC}-\ref{Fig:SG} we show results for $J_2=0.55$. Figs.\ref{Fig:SLSC}-\ref{Fig:SLDSG} summarize results for $J_2=0.5$ to facilitate a comparison.

\begin{figure*}
  \includegraphics[width=1.0\linewidth]{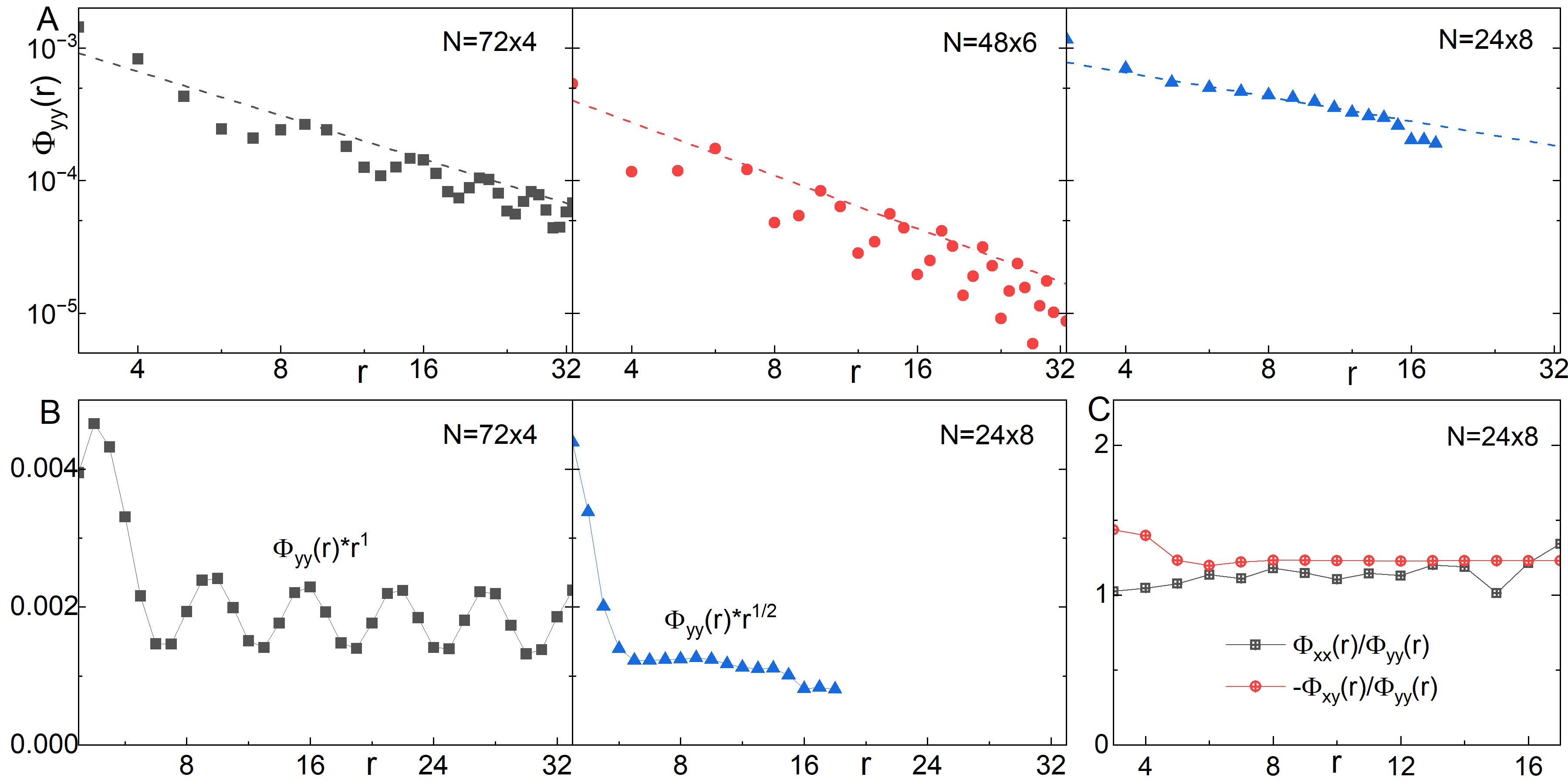}
\caption{(Color online) Superconducting correlations, $\Phi_{yy}(r)$,  versus  distance, $r$, between two Cooper pairs in the $\hat{x}$ direction for $J_2=0.55$ and $\delta=1/12$: (A) On a log-log where the dashed lines denote power-law fits. (B) On a linear plot with ordinate is rescaled as $\Phi_{yy}(r) r^1$ for $W=4$ and $\Phi_{yy}(r) r^{1/2}$ for $W=8$. (C) Ratios of $\Phi_{xx}/\Phi_{yy}$ and -$\Phi_{xy}/\Phi_{yy}$ on a $N=24\times 8$ cylinder.} \label{Fig:SC}
\end{figure*}

\section{Results}\label{Sec:Results}
\subsection{Superconducting pair-field correlations }\label{Sec:SC}
To probe superconductivity, we have calculated  the equal-time spin-singlet SC 
pair-field correlation function 
\begin{eqnarray}
\Phi_{\alpha\beta}(r)=\langle\Delta^{\dagger}_{\alpha}(x_0,y)\Delta_{\beta}(x_0+r,y)\rangle. \label{Eq:SC}
\end{eqnarray}
Here $\Delta^{\dagger}_{\alpha}(x,y)=\frac{1}{\sqrt{2}}[\hat{c}^{\dagger}_{(x,y),\uparrow}\hat{c}^{\dagger}_{(x,y)+\alpha,\downarrow}+\hat{c}^{\dagger}_{(x,y)+\alpha,\uparrow}\hat{c}^{\dagger}_{(x,y),\downarrow}]$ is the spin-singlet pair creation operator on a bond in the $\alpha=\hat{x}$ or $\hat{y}$ direction, and ($x_0,y$) is a reference site taken as $x_0\sim L/4$ and $r$ is the  displacement between two bonds in the $\hat x$ direction.

In Fig.\ref{Fig:SC}A we show $\Phi_{yy}(r)$ versus $r$ on a log-log scale for all values of $W$ with $J_2=0.55$ and $\delta=1/12$. The dashed lines in the figure represent a power law decay with the power chosen for each $W$ to provide a best-fit (for $W=4$ and 6) thru the locus of peak heights of the decaying oscillations. In Fig.\ref{Fig:SC}B we show the same data for $W=4$ and 8 as in panel A, now on a linear-linear plot, but re-scaled by a simple power law such that the vertical axis for $W=4$ is $r^1\Phi_{yy}(r)$ and is $r^{1/2}\ \Phi_{yy}(r)$ for $W=8$. Fig.\ref{Fig:SC}C shows the ratio of $\Phi_{xx}(r)/\Phi_{yy}(r)$ and $-\Phi_{xy}(r)/\Phi_{yy}(r)$ for $W=8$ on a linear-linear plot. Data for other values of $\delta$ and $J_2$, as a function of the number of kept states $m$, are presented in Fig.\ref{Fig:SLSC} and Appendix \ref{SM:SC}

\begin{figure*}
  \includegraphics[width=1.0\linewidth]{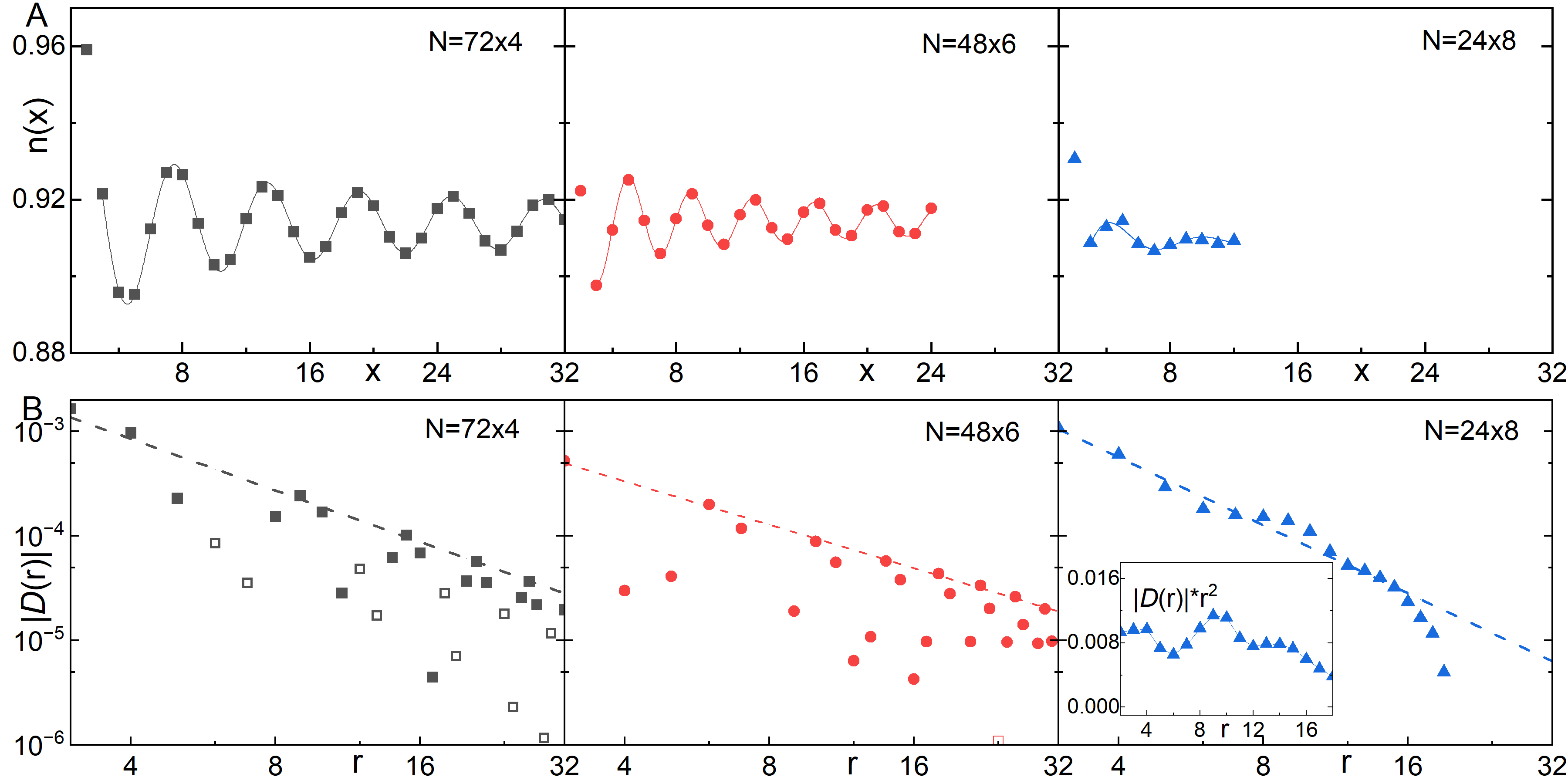}
\caption{(Color online) Charge density correlations for $J_2=0.55$ and $\delta=1/12$. (A) Charge density profiles $n(x)$ on $W=4,6,8$ cylinders where $x$ is the rung coordinate. (B) Log-log plot of the rung charge density correlations $|D(r)|$ versus the distance $r$ between two rungs in the $\hat{x}$ direction. The filled/open symbols represent the sign of $D(r)$ being $-/+$, respectively. Inset: Re-scaled $|D(r)|r^2$ for the $W=8$ cylinder.}\label{Fig:CDW}
\end{figure*}

Given that we can access relatively large values of $L$, it is reasonable to analyze these results in the context of the expected behavior for fixed $W$ in the $L \to \infty$ limit, where the asymptotic behavior is ultimately that of an effective 1D quantum field theory.  Since continuous symmetries cannot be broken in 1D, quasi-long-range  order is the strongest indication of SC order that can be expected:
\begin{eqnarray}
\Phi_
{ab}(r)\sim  r^{-K_{sc}}.\label{Eq:Ksc}
\end{eqnarray}
Given the fact that at large $r$, after re-scaling, the $W=8$ result, and the mean of the oscillatory part of $W=4$ result, are approximately independent of $r$ is evidence that the data in Fig.\ref{Fig:SC}B is  consistent with Eq.(\ref{Eq:Ksc}) with exponents $K_{sc}\approx 1$ for $W=4$ and $K_{sc}\approx 1/2$ for $W=8$. (Similar scaling using the exponents given in Table\ref{Table:Sum}, not shown, works for W=6 as well.) The ``best fit'' values of these exponents obtained in Fig.\ref{Fig:SC}A are given in Table \ref{Table:Sum}, namely, $K_{\rm sc}(W=4)\approx 1.1$, $K_{\rm sc}(W=6)\approx 1.3$, and $K_{\rm sc}(W=8)\approx 0.6$. (The meaning and uncertainties of the ``best fit'' are discussed in Sec. \ref{Sec:SumRes}.) 

One subtlety that is apparent in the data for $W=4$ and $6$, but not present for $W=8$, is the presence of spatial modulations of $\Phi$ with the same ordering vector ${\bf Q}$ as the CDW correlations (see Fig.\ref{Fig:CDW}A below). The nearly constant amplitude of the modulation in Fig.\ref{Fig:SC}B implies the amplitude of the SC pair density modulation decays with nearly the same exponent as does the uniform SC order. In principle this could signify the presence of significant pair-denisty-wave (PDW) correlations. However, this would require a highly unlikely accidental degeneracy of the decay exponent for the uniform SC and PDW orders. Instead, we believe this behavior  reflects the fact that our calculations are carried out at finite $L$, and is a consequence of the pinning of the CDW fluctuations by the boundary, as are the charge density oscillations shown in Fig.\ref{Fig:CDW}A. We shall elaborate on this point in  Appendix \ref{SM:SC-PDW}.

Significantly, within numerical accuracy, $\Phi_{yy}(r)$, $\Phi_{xx}(r)$ and $-\Phi_{xy}(r)$ are all characterized by the same decay exponent $K_{\rm sc}$, as is shown for $W=8$ in Fig.\ref{Fig:SC}C. Invoking the expected asymptotic Lorenz symmetry of 1D systems, one can infer that the SC susceptibility should diverge as $\chi_{\rm sc}\sim T^{-(2-K_{\rm sc})}$ as $T\rightarrow 0$, and hence a smaller value of $K_{\rm sc}$ implies a stronger divergence.

\subsection{Charge density wave correlations}\label{Sec:CDW}
For $W=4$ and $6$, we observe a tendency to form charge stripes in the lightly doped cylinders with $J_2=0.55$ that is similar to that reported in earlier work at $J_2=0.5$\cite{Jiang2021}. In comparison, all signatures of CDW ordering are much weaker for $W=8$.

To measure the CDW order, we define the rung charge density $n(x)=W^{-1}\sum_{y=1}^{W}\langle \hat{n}(x,y)\rangle$. Fig.\ref{Fig:CDW}A shows examples of $n(x)$ on $W=4,6,8$ cylinders with $J_2=0.55$ at $\delta=1/12$, where $x$ is the distance from one end of the cylinder up to a maximum value $x=L/2$. The charge density oscillations have a period $\lambda$ that is consistent with $\delta\ W\ \lambda=2$, namely, there are two holes per unit cell. When divided by $W$ this amounts to ``half-filled stripes" for the case $W=4$ and ``one third filled strpies'' for $W=6$. For $W=8$, the oscillatory component of $n(x)$ is much weaker, which combined with the relatively small range of accessible $x$ ($\leq L/2$) makes extracting the period less reliable.

\begin{figure*}
  \includegraphics[width=1.0\linewidth]{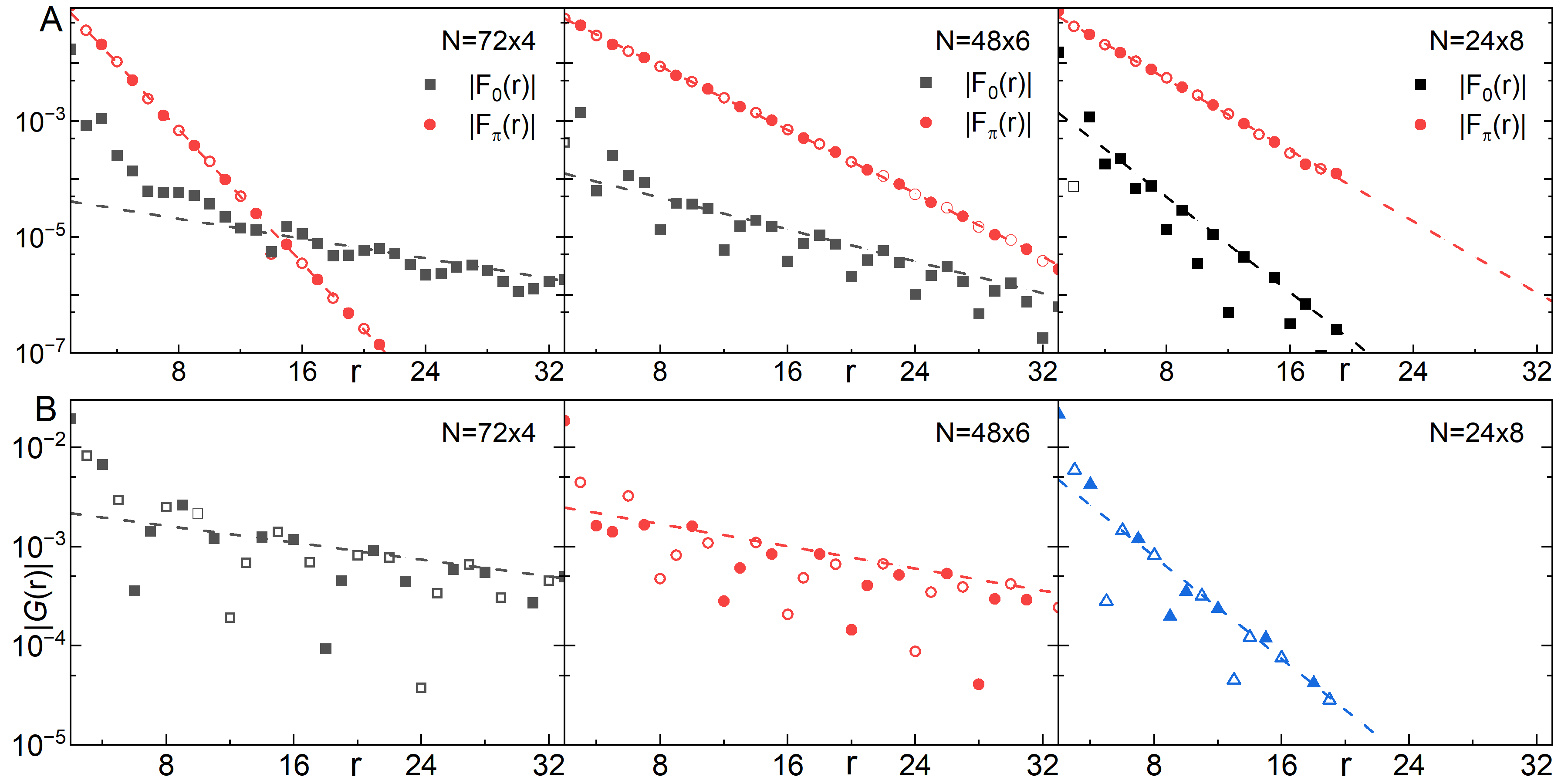}
\caption{(Color online) Spin-spin and single-particle correlations for $J_2=0.55$ and $\delta=1/12$. (A) Log-linear plot of $|F_0(r)|$ and $|F_\pi(r)|$ for the $W=4,6,8$ cylinders. Dashed lines denote exponential fits $|F_0(r)|\sim e^{-r/\xi_F}$ and $|F_\pi(r)|\sim e^{-r/\xi_{AF}}$ with correlation length $\xi_F$ and $\xi_{AF}$. (B) Log-linear plot of $|G(r)|$ for the $W=4,6,8$ cylinders. Dashed lines denote exponential fit $|G(r)|\sim e^{-r/\xi_G}$ with single-particle correlation length $\xi_G$. The filled/open symbols represent the sign of $F_0(r)$, $F_\pi(r)$ and $G(r)$ being $-/+$, respectively.}\label{Fig:SG}
\end{figure*}

The oscillations in Fig.\ref{Fig:CDW}A are presumably a finite $L$ effect - reflecting the pinning of the CDW fluctuations by the cylinder ends.  At long distances, the spatial decay of the CDW correlations associated with such ``generalized Friedel oscillations'' are governed\cite{White2002} by the CDW 
Luttinger exponent, $K_c$, as
\begin{eqnarray}
n(x)-n_0 \sim A_Q\ast {\rm cos}(Qx + \phi) x^{-K_c/2}.\label{Eq:KcNx}
\end{eqnarray}
Here $A_Q$ and $\phi$ are, respectively, a non-universal amplitude and phase shift, $n_0=1-\delta$ is the mean electron density, and $Q=2\pi/\lambda$ is the dominant charge density ordering wave-vector.  We find this formula works well for the $W=4$ and $W=6$ cylinders with $J_2=0.55$, as it did for $J_2=0.5$\cite{Jiang2021}. {The ``best fit''} value of the Luttinger exponent from the decay of $n(x)$ yield $K_c(W=4)\approx 1.2$, $K_c(W=6)\approx 1.4$, $K_c(W=8)\geq 2$, respectively. However, for $W=8$ cylinders the charge oscillations are much weaker,  and the range of $|r|<L$ more restricted, so obtaining a value of $K_c$ in this way is subject to large uncertainty. 

A value of the exponent $K_c$ can also (independently) be extracted from the charge density-density fluctuation correlation function, defined as 
\begin{align}
\label{Dofr}
&D(r)= \\
&\langle\left[\hat{n}(x_0,y)-\langle \hat{n}(x_0,y)\rangle\right] 
\left[\hat{n}(x_0+r,y)-\langle \hat{n}(x_0+r,y)\rangle\right]\rangle.
\nonumber
\end{align}
Here ($x_0,y$) is a reference site and $r$ is the distance between two sites in the $\hat{x}$ direction and $x_0\sim L/4$. Fig.\ref{Fig:CDW}B shows $D(r)$ on $W=4,6,8$ cylinders at $\delta=1/12$. Based on field theoretic (i.e. bosonization) considerations we expect that 
\begin{eqnarray}
D(r)\sim A_0^\prime\ |r|^{-K_{c0}} + A_Q^\prime\ \cos(Qr+\phi^\prime)|r|^{-K_c} + \ldots\label{Eq:Kc}
\end{eqnarray}
where again $Q$ is the wave-vector of the dominant CDW correlations, $A_0^\prime$, $A_Q^\prime$, and $\phi^\prime$ are non-universal constants, the $\ldots$ represent oscillations at other wave-vectors - harmonics of $Q$ or in cases where there is more than one gapless mode (presumably not relevant in present circumstances) at subdominant CDW ordering wave-vectors.  Here $K_c$ is the same Luttinger exponent already discussed while on general grounds one expects  $K_{c0}=2$ in the presence of any sound-like compressional mode. Values of $K_c$ corresponding to the dashed line ``best fit'' to the data in Fig.\ref{Fig:CDW}B yield $K_c(W=4)\approx 1.6$, $K_c(W=6)\approx 1.4$, and $K_c(W=8)\geq 2$. Note that while $K_c$ extracted from $D(r)$ for $W=4$ and 6 are slightly different from those extracted from $n(x)$ \cite{Jiang2021}, they are qualitatively similar.

Extracting a value of $K_c$ from $D(r)$ is particularly difficult whenever $K_c > K_{c0}=2$, as in this case the CDW correlations contribute a subdominant piece.  Indeed, an oscillating piece of $D(r)$ at the longest distances is not clearly identifiable in  our data for $W=8$.  We thus consider it likely that the value of $K_c\approx 2$ obtained from the dashed line fit to the data with $W=8$  corresponds to the value of $K_{c0}$, consistent with inferred value of $K_c(W=8)>2$ obtained from the fit to $n(x)$. This holds true for both $J_2=0.55$ and $J_2=0.5$ (Fig.4), which demonstrates the clear suppression of charge order on wider systems. 

It is worth emphasizing that on the basis of the fact that $K_c<2$ for both $W=4$ and $W=6$ cylinders, the same field-theoretic analysis implies that the CDW susceptibility diverges $\chi_c\sim T^{-(2-K_c)}$ as $T\rightarrow 0$. In contrast, the charge susceptibility $\chi_c$ on $W=8$ cylinders 
does not diverge even at $T=0$ since $K_c(W=8)\geq 2$. More results on the CDW correlations are given in Appendix \ref{SM:CDW}.

\begin{figure*}[!th]
\centering
  \includegraphics[width=1.0\linewidth]{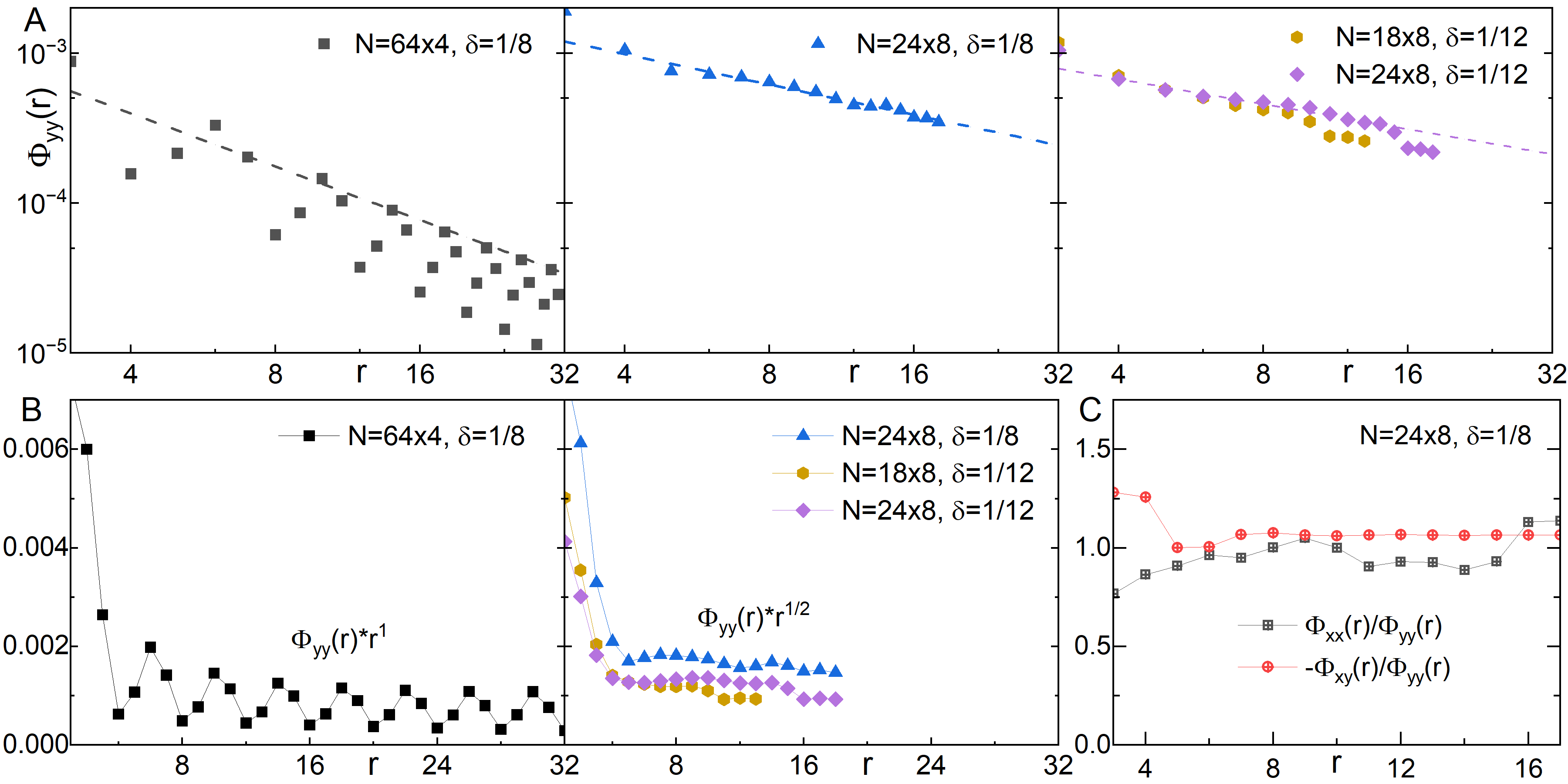}
\caption{(Color online) Superconducting correlations $\Phi_{yy}(r)$, versus distance $r$ between two Cooper pairs in the $\hat{x}$ direction for $J_2=0.50$, $\delta=1/8$ and $\delta=1/12$: (A) On a log-log plot where the dashed lines denote power-law fits. (B) On a linear-plot with ordinate rescaled as $\Phi_{yy}(r)r^1$ for $W=4$ and $\Phi_{yy}(r)r^{1/2}$ for $W=8$. (C) Ratios of $\Phi_{xx}/\Phi_{yy}$ and $-\Phi_{xy}/\Phi_{yy}$ on a $N=24\times 8$ cylinder at $\delta=1/8$.} \label{Fig:SLSC}
\end{figure*}

\subsection{Spin-spin and single-particle correlations }\label{Sec:SG}
To describe the magnetic properties of the ground state, we calculate the equal-time spin-spin correlation functions defined as%
\begin{eqnarray}\label{Eq:SpinCor}
F_q(r)=W^{-2} \sum_{y,y'}e^{iq(y-y')}\langle \vec{S}_{x_0,y}\cdot \vec{S}_{x_0+r,y'}\rangle,
\end{eqnarray}
where $x_0\sim L/4$ is a reference rung and $q=2\pi m/W$ is a transverse momentum with  $0\leq m < W$ is an integer. Fig.\ref{Fig:SG}A shows $|F_0(r)|$ and $|F_\pi(r)|$ on a log-linear scale for $J_2=0.55$ and $\delta=1/12$, and Fig.\ref{Fig:SLDSG} B shows 
the same quantities for $J_2=0.5$ and $\delta=1/12$ and $1/8$. The corresponding correlation functions for other values of $q$ are generally much smaller and more rapidly falling at large $r$.  

In the figures, the sign of $F_q(r)$ is indicated by closed (positive) and open (negative) symbols.  Thus, from the pattern of open and closed symbols, it can be seen that  $F_\pi$ corresponds to locally Neel AF order (i.e. it has an ordering vector of roughly $(\pi,\pi)$), while $F_0$ is dominated by locally ferromagnetic correlations.  We thus identify two spin correlation lengths, $\xi_{\rm AF}$ and $\xi_F$, from the decay of these two quantities. The fact that $\xi_{\rm AF}$ is always less than $W$ and shows no clear tendency to increase with $W$ (see Table \ref{Table:Sum}) suggests that in the 2D limit it is finite. This is consistent with the expected behavior of a quantum paramagnet with short-range antiferromagnetic correlations. Interestingly Fig.\ref{Fig:SG}A and Fig.\ref{Fig:SLDSG} B also show that ferromagnetic correlations arise which, for the case of $W=4$ have rather long correlations lengths, with $\xi_F \gg \xi_{AF}$!  The reason for this is currently unclear to us.  We have, however, checked that these correlations are induced by doping. Except for $W=4$, $\xi_F$ is close to $\xi_{\rm AF}$ (see Table \ref{Table:Sum}), suggesting that ferromagnetic correlations are also short-ranged in the 2D limit.

We have also calculated the equal-time single-particle Green function, defined as%
\begin{eqnarray}\label{Eq:CC}
  G(r)=\langle c^{\dagger}_{(x_0,y),\sigma} c_{(x_0+r,y),\sigma}\rangle.
\end{eqnarray}
Fig.\ref{Fig:SG}B shows $G(r)$ for $W=4,6,8$ cylinders with $J_2=0.55$ at $\delta=1/12$. At long distances, $G(r)$ is also consistent with an exponential decay $G(r)\sim e^{-r/\xi_G}$. The extracted correlation lengths, $\xi_G$, are  given in Table \ref{Table:Sum}. $\xi_G$ also decreases with increasing $W$, suggesting that at this doping $\xi_G$ is finite in two dimensions. Again, the same behavior is observed for $J_2=0.5$.

\begin{figure}
  \includegraphics[width=\linewidth]{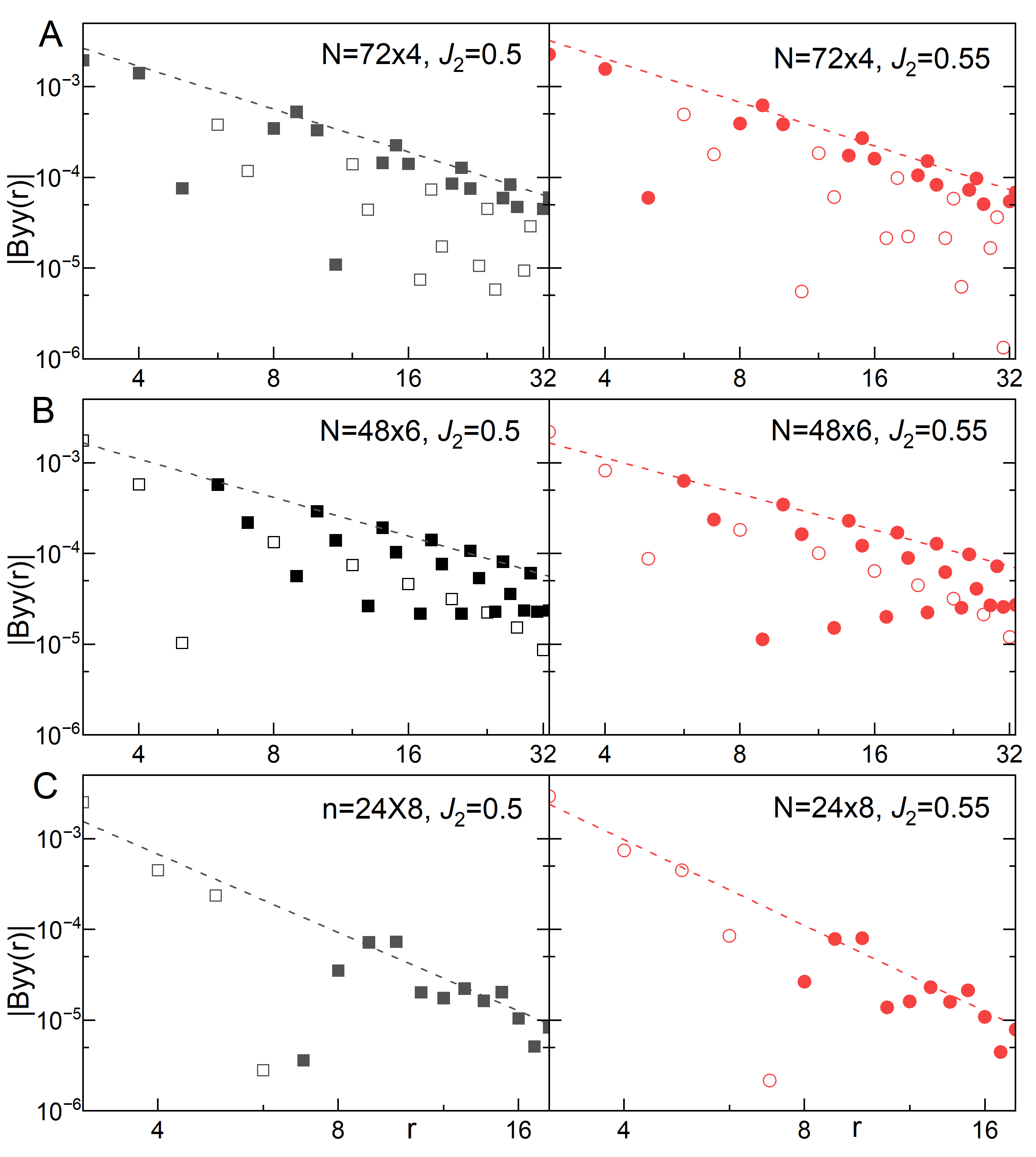}
\caption{(Color online) Log-log plots of $|B_{yy}(r)|$ on (A) $N=72\times 4$, (B) $N=48\times 6$ and (C) $N=24\times 8$ cylinders at $\delta=1/12$ for $J_2=0.5$ and $J_2=0.55$. Here $r$ is the distance between two bonds in the $\hat{x}$ direction. Here filled/open symbols represent the sign of $B_{yy}(r)$ being $-/+$, respectively.} \label{Fig:Dimer}
\end{figure}

\subsection{The dimer-dimer correlations}\label{Sec:Dimer}%
We have also calculated the dimer-dimer correlation function which is defined as%
\begin{eqnarray}
B_{\alpha\beta}(r)&=&\langle B_\alpha(x_0,y) B_\beta(x_0+r,y)\rangle \nonumber \\%
&-& \langle B_\alpha(x_0,y)\rangle \langle B_\beta(x_0+r,y)\rangle.\label{Eqs:Dimer}
\end{eqnarray}
Here $B_\alpha(x,y)=\vec{S}_{(x,y)}\cdot \vec{S}_{(x,y)+\alpha}$ is the (spin) dimer operator on bond $\alpha=\hat{x}$ or $\hat{y}$, and $(x_0,y)$ is a reference bond taken as $x_0\sim L/4$ and $r$ is the distance between bonds in the $\hat{x}$ direction.

 Fig.\ref{Fig:Dimer} shows log-log plots of the dimer-dimer correlations $|B_{yy}(r)|$ for $W=4,6,8$ cylinders at $\delta=1/12$ with both $J_2=0.5$ and $J_2=0.55$. For $W=4,6$ cylinders, the dimer correlation has same oscillatory period as the CDW.  All $B_{yy}(r)$ shown in the figure are in the limit $m=\infty$ which are obtained using a second-order polynomial function to fit the four data points associated with the largest number of kept states.  From the figure we can see that  there are only small quantitative differences in the behavior of $B_{yy}(r)$ between $J_2=0.5$ and $J_2=0.55$. The extracted exponent $K_{\rm dimer}$ is provided in Table \ref{Table:Sum}.

\begin{figure*}[!th]
\centering
  \includegraphics[width=1.0\linewidth]{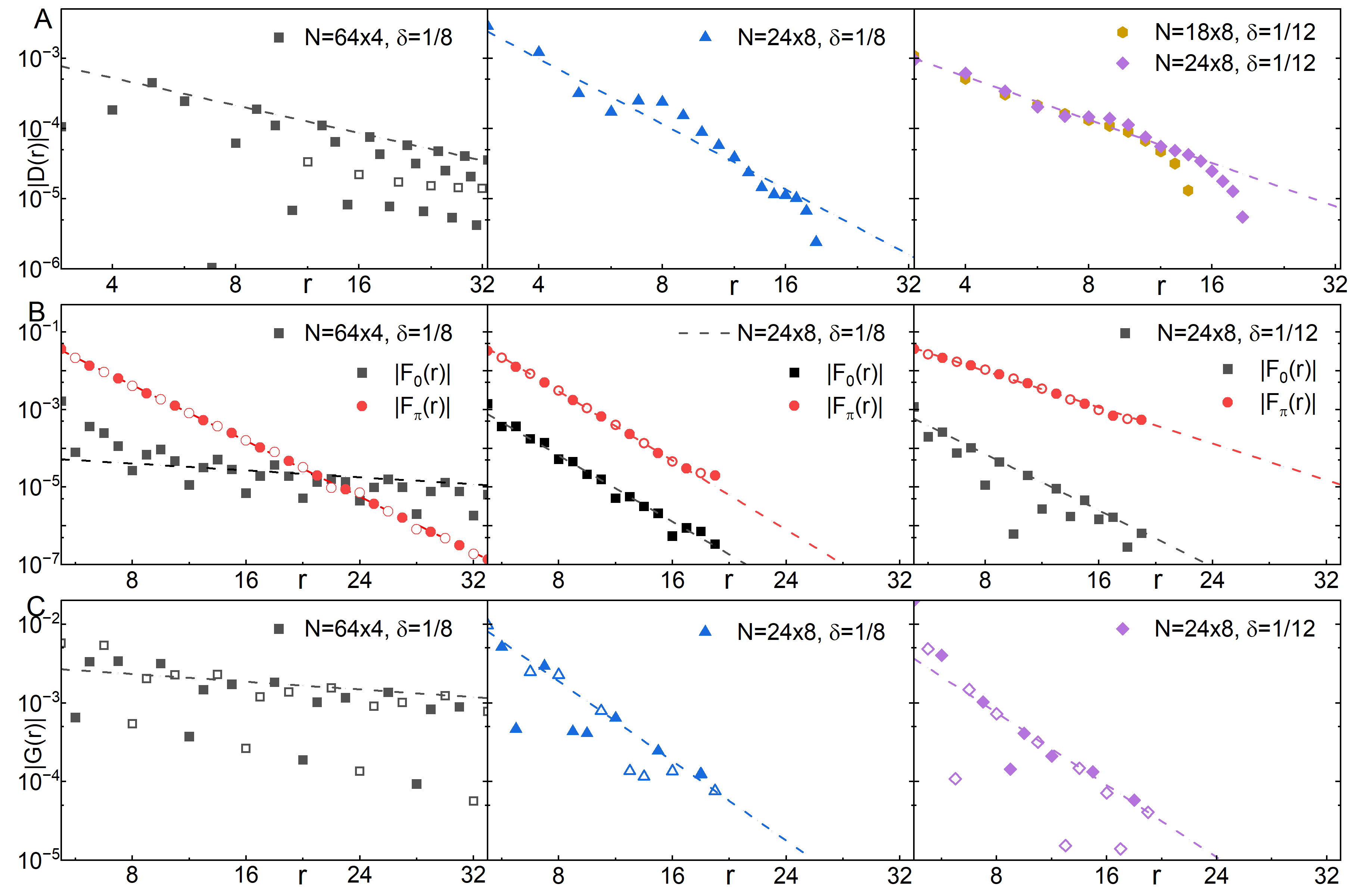}
\caption{(Color online) Other correlations for $J_2=0.50$ at $\delta=1/8$ and $\delta=1/12$. (A) Log-log plot of the rung charge density correlations $|D(r)|$ where dashed lines denote power-law fits. (B,C) The Log-linear plot of $|F_0(r)|$, $|F_\pi(r)|$ and $|G(r)|$ where dashed lines denote exponential fits, between two sites in the $\hat{x}$ direction. Here the definition of r is the same as in the previous figures. The filled/open symbols represent the sign of $D(r)$, $F_0(r)$, $F_\pi(r)$ and $G(r)$ being $-/+$, respectively.} \label{Fig:SLDSG}
\end{figure*}

\subsection{Results for $J_2=0.5$}\label{Sec:J05}
We have also studied the doped model with $J_2=0.5$ on cylinders with $W=8$, extending earlier work\cite{Jiang2021} with the same $J_2$ but  smaller $W$. Representative results are shown in Fig.\ref{Fig:SLSC} and Fig.\ref{Fig:SLDSG} which include the various correlation functions, $\Phi_{yy}(r)$, $D(r)$, $F_0(r)$, $F_\pi(r)$, and $G(r)$ for $\delta=1/12$ and $\delta=1/8$. We find that the qualitative behaviors for all these correlation functions including  the $W$ dependence of exponents $K_{\rm sc}$ and $K_c$, and the correlation lengths $\xi_{AF}$, $\xi_F$, and $\xi_G$, are very similar to those  at $J_2=0.55$. For both cases, we observed quasi-long-range ordered SC. Moreover, the isotropy of $\Phi_{yy}/\Phi_{xx}~{\rm and}~-\Phi_{xy}/\Phi_{xx}$, the dependence of  $K_{\rm sc}$ on $W$, and the lack of other competing orders, suggest that what is observed could be the manifestation of a 2D long-range ordered superconductivity (without CDW order) on finite cylinders.

\subsection{Summary of results}\label{Sec:SumRes}
In Table.\ref{Table:Sum}, we summarize our results shown in the main text, including the system sizes, doping concentration and coupling parameters, the extracted Luttinger exponent $K_{\rm sc}$ for SC correlations, $K_c$ for charge density correlations, $K_{\rm dimer}$ for dimer-dimer correlations, as well as extracted correlation lengths for both spin-spin and single-particle correlations. Note that we have not included error bars. This is  because errors can originate from multiple sources, e.g., from the finite $m$ extrapolation, the effects of finite $L$, and the uncertainty in the assumed fitting functions, etc. Although, once a particular fitting function is assumed, we can obtain error bars associated with the range of parameters consistent with a fit, we refrain from doing so, as the resulting error bars tend to be relatively small, and thus could produce a false impression concerning the certainty of the inferred exponents and correlation lengths. In the same spirit, when we give a ``best fit'' value in the main text, it means we have committed to a fitting function and ignored the above uncertainties.

\begin{table*}[tb]
\caption{Summary of results shown in Figs.\ref{Fig:SC}-\ref{Fig:SLDSG} and in the Appendices for the given values of $J_2/J_1$ and $\delta$. $K_{sc}$ is the SC Luttinger exponent extracted from a ``best fit'' to the long-distance decay of $\Phi_{yy}(r)$. $K_c$ is the CDW Luttinger exponent extracted by fitting two different quantities -  the charge density profile $n(x)$ and the decay of the density-density correlation function $D(r)$. $K_{\rm dimer}$ is the dimer power-law decay exponent extracted from a ``best fit" to the dimer-dimer correlation function $B_{yy}(r)$. $\xi_{\rm AF},\xi_F$ and $\xi_G$ are the spin-spin and single-particle correlation lengths. Note that  there are two spin-spin correlation lengths: $\xi_{\rm AF}$ is associated with the decay of locally commensurate, $Q=(\pi,\pi)$ correlations while $\xi_{F}$ is associated with locally ferromagnetic correlations.} 
\centering 
\begin{tabular}{c | c | c | c | c | c | c | c | c | c  } 
\hline\hline 
 $\#$ of sites & $J_2/J_1$ & $\delta$ & $K_{sc}$ & $n(x)\Rightarrow K_c$ &  $D(r)\Rightarrow K_c$ & $K_{\rm dimer}$ & $\xi_{\rm AF}$ & $\xi_{\rm F}$ & $\xi_G$ \\ [.5ex] 
\hline 
 $N=72\times 4$ & 0.55 & 1/12 & $1.1$ & $1.2$ & $1.6$ & $1.6$ & $1.9$ & $10$ & $21$ \\
 $N=48\times 6$ & 0.55 & 1/12 & $1.3$ & $1.4$ & $1.4$ & $1.3$ &  $2.2$ & $8.7$ & $16$ \\
 $N=24\times 8$ & 0.55 & 1/12 & $0.6$ & $\ge 2$ & $\ge 2$ & $3.1$ & $2.8$ & $2.1$ & $3.4$ \\
 $N=72\times 4$ & 0.50 & 1/12 & $1.1$ & $1.3$ & $1.7$ & $1.6$ &  $1.8$ & $13$ & $30$ \\
 $N=48\times 6$ & 0.50 & 1/12 & $1.3$ & $1.5$ & $1.4$ & $1.4$ &  $3.7$ & $6.5$ & $21$\\
 $N=24\times 8$ & 0.50 & 1/12 & $0.6$ & $\ge 2$ & $\ge 2$ & $2.9$ & $3.7$ & $2.4$ & $3.8$ \\
 $N=64\times 4$ & 0.50 & 1/8 & $1.2$ & $1.4$ & $1.3$ & $1.4$ &  $2.4$ & $20$ & $36$ \\
 $N=24\times 8$ & 0.50 & 1/8 & $0.7$ & $\ge 2$ & $\ge 2$ & $3.3$ & $2.0$ & $2.1$ & $3.4$ \\
\hline\hline 
\end{tabular}
\label{Table:Sum}
\end{table*}

\section{Summary and Conclusions}\label{Sec:Summary}
We have studied the $t$-$J$ model with $J_2/J_1=0.5$ and $0.55$. According to 
Refs.\cite{Gong2014,Wang2018,Liu2022SciBul,Liu2022PRX} for these values of $J_2/J_1$ the undoped insulating ``parent state'' at half-filling is in the QSL and VBC phases, respectively. While the precise nature of the phases of the undoped, 2D insulator is still under debate, the preponderance of the evidence suggests  that for these values of $J_2/J_1$ they are paramagnets with at most weak VBC order. The behaviors found in the present DMRG studies on moderately lightly doped cylinders with $W\le 8$ motivate us to propose  that the SC state we have found survives in the 2D ($W\to\infty$) limit.

To corroborate this conclusion, it is illuminating to compare our results to the expected behavior of a 2D superconductor restricted to infinitely long cylinders ($L \to \infty$) with large but non-infinite $W$ (See Ref.\cite{gannot}). Since the system is ultimately one dimensional for any finite $W$, quantum phase fluctuations ensure that only SC quasi-long-range order is possible, and for large $W$ one expects the power-law decay exponent $K_{sc} \sim 1/W$. Moreover, the SC order parameter in 2D determines the cylinder SC correlations via $\Phi_{\alpha\beta}(r)\sim \Delta_\alpha^\star \Delta_\beta \ r^{-K_{sc}}$  where $\Delta_\alpha$ is the 2D expectation value of the SC order parameter on neighboring sites in the $\alpha=x,$ $y$ directions. The fact that $\Phi_{yy}(r)/\Phi_{xx}(r) \approx -\Phi_{xy}(r)/\Phi_{xx}(r) \approx 1$ at large $r$ is consistent with the expectation that the SC state is $d$-wave in the 2D limit. Although there is considerable uncertainty in the accuracy with which they can be determined, the inferred values of $K_{sc}$ for $W=4$ and 8 are roughly consistent with the expected scaling behavior, i.e. $K_{sc} (W=8)/K_{sc}(W=4)\approx 1/2$. Note, however, that the inferred value for $W=6$ is notably larger than would be expected on this basis. Indeed, there is no compelling reason to expect the large $W$ scaling analysis to be applicable down to $W$'s as small as 4 or even 8.

Turning to other correlations, if there is no CDW order and no gapless Fermi surface in the 2D limit, one expects the CDW correlation function $D(r)$ to fall with a power law $K_{c} \sim 1/K_{\rm sc}$ on finite cylinders. If $K_c>2$,  $D(r)$ is expected to be dominated by the long-wave-length acoustic modes of the electron density, which always fall as $r^{-2}$.  This is also roughly consistent with the results for $W=8$.  The estimated  dimer-dimer correlations also fall with a decay exponent $K_{\rm dimer}\approx K_c >K_{\rm sc}$. As to the spin-spin correlation function, the existence of a relatively short (compared with W) spin correlation length and a correspondingly robust spin-gap in all cases suggests the SC state is a nodeless d-wave in the 2D limit. This notion is supported by the exponential decay of the equal-time single-particle Greens function. The lack of nodal quasiparticles 
is probably best viewed from a strong coupling limit in which the Cooper pairs correspond to real-space valence bonds - quantum dimers.\cite{rokhsarandme} In this case, the  d-wave symmetry of the pair wavefunction does not necessarily imply gap nodes. However, we should add that the above "Occam-razor" type interpretation of the data does not prove that for even larger $W$ a new trend will not emerge.

\indent{In} our physical picture, the existence of singlet pair correlations is not the only requirement 
for doping-induced superconductivity.  Another requirement is that the hopping of the doped holes must generate significant superfluid stiffness. An example of such hole mobility induced superconductivity in a VBC is given in Ref.\cite{Li2023}. However, it is simultaneously important that hole hopping does not overly disrupt the singlet correlations of the ``parent'' state. When $t_2 <0$,  the single hole kinetic energy is minimized in a ferromagnetic background (a generalized Nagaoka's theorem\cite{tasaki}), which implies a strong tendency to destroy local singlet correlations.  It has been argued\cite{kanelee} that this leads to a large mass renormalization of the doped holes.  Conversely, in Ref. \cite{kyungsu} it was shown that on a ``triangular cactus lattice''the kinetic energy of a single hole, with negative $t_1$, is minimized in a resonating valence bond liquid, i.e. the hole kinetic energy actually stabilizes local singlet formation.  A related suggestion presented in Ref.\cite{Martins2001} is that $t_2/t_1$ positive (negative) causes constructive (destructive) interference when a pair of holes move in a spin-singlet background.

Given how hard it has proven to find any material that demonstrably is a QSL without doping\cite{broholmreview}, looking for dopable spin liquids may not be the most practical strategy to identify new and interesting superconducting materials.  In this context, the
fact that an essentially identical SC state can be reached by 
doping what is likely a weak VBC may offer an additional clue in the search for new superconductors.

 It is important to point out that while the values of $\delta$ we have studied are relatively small, we have not directly addressed the behavior of the system in the limit as $\delta \to 0$. What doping concentration is ``sufficient'' depends on the state at half-filling. For example, the critical doping for inducing SC in a QSL will likely be considerably smaller (and could even vanish\cite{rokhsarandme,IoffeLarkin}) than that for a VBC.

{\it Acknowledgments:} We thank Zheng-Yu Weng, Tao Xiang, Ashvin Vishwanath and Hong Yao for helpful discussions. (H-C.J. and S.A.K.) was supported by the Department of Energy (DOE), Office of Sciences, Basic Energy Sciences, Materials Sciences and Engineering Division, under Contract No. DE-AC02-76SF00515. (D-H. L.) was supported in part by the U. S. Department of Energy, Office of Science, Office of Basic Energy Sciences, Materials Sciences and Engineering Division under Contract No. DE-AC02-05-CH11231 (Theory of Materials program KC2301).

\appendix 
\renewcommand{\thefigure}{A\arabic{figure}}
\setcounter{figure}{0}
\renewcommand{\theequation}{A\arabic{equation}}
\setcounter{equation}{0}
\setcounter{table}{0}
\renewcommand{\thetable}{A\arabic{table}}
\setcounter{section}{0}

\section{Superconducting correlations}\label{SM:SC}%
\begin{figure}
  \includegraphics[width=\linewidth]{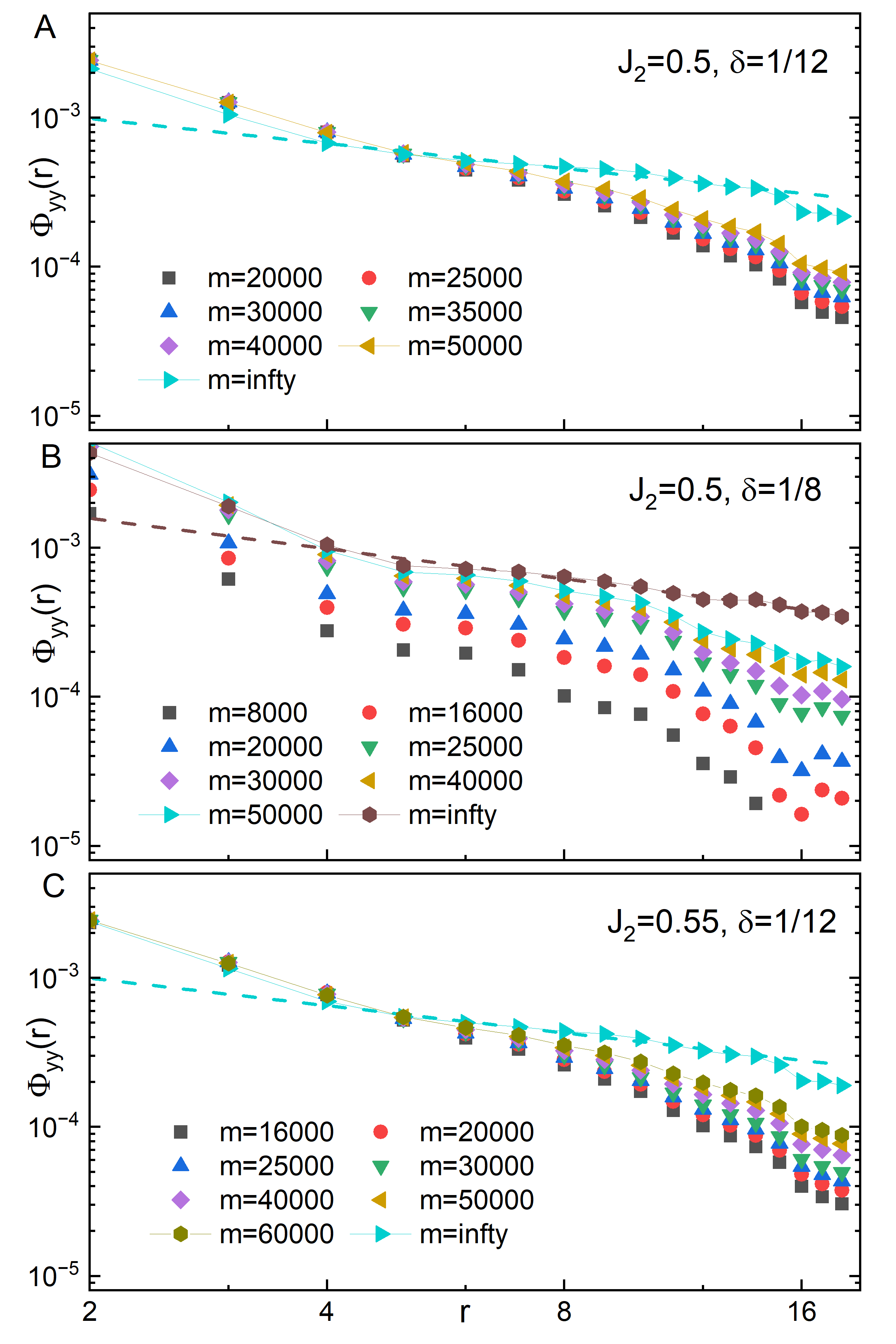}
\caption{(Color online) Superconducting correlation $\Phi_{yy}(r)$ in log-log scales on $N=24\times 8$ cylinders at (A) $\delta=1/12$ with $J_2=0.5$, (B) $\delta=1/8$ with $J_2=0.5$ and (C) $\delta=1/12$ with $J_2=0.55$, by keeping $m$number of states and its extrapolation to the limit $m=\infty$. The dashed lines denote the power-law fits $\Phi_{yy}(r)\sim r^{-K_{sc}}$.} \label{FigA:SC}
\end{figure}

Fig. \ref{FigA:SC} shows the log-log plots of the SC pair-field correlations $\Phi_{yy}(r)$ for $N=24\times 8$ for different numbers of kept states $m$. The top two panels are for $J_2=0.5, \delta=1/12$ and $J_2=0.5,\delta=1/8$; the bottom panel is for $J_2=0.55$, $\delta=1/12$. The extrapolated $\Phi_{yy}(r)$ in the $m\rightarrow\infty$ limit is obtained using a second-order polynomial function fitting the four data points with the largest $m$. A power-law fit to  the SC correlations of the form $\Phi_{yy}(r)\propto r^{-K_{\rm sc}}$ is indicated by the dashed lines. To exclude the short-distance behavior and the boundary effects due to finite $L$, the data points with the smallest and the largest $r$  were omitted in obtaining this fit. For $J_2=0.5$, $\delta=1/12$ the ``best-fit'' exponent is $K_{\rm sc}\approx 0.56$ and for $J_2=0.5$, $\delta=1/8$, $K_{\rm sc}\approx 0.67$. For $J_2=0.55$, $\delta=1/12$, the ``best-fit'' exponent is $K_{\rm sc}\approx 0.61$. In addition to the spin-singlet Cooper pair correlations, we have also calculated the spin-triplet Cooper pair correlations. However, these are much weaker, suggesting that spin-triplet superconductivity is unlikely.

\begin{figure}
  \includegraphics[width=\linewidth]{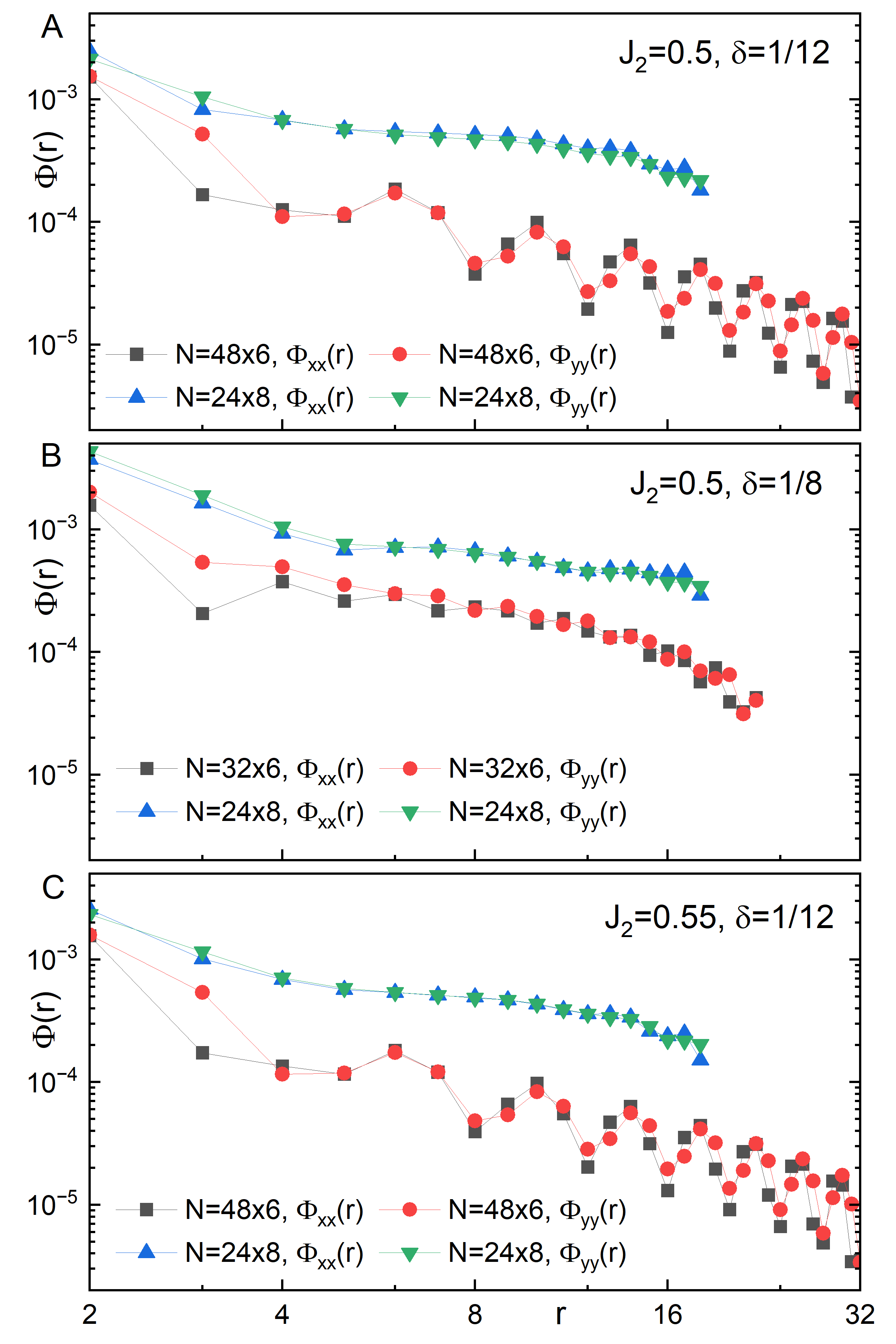}
\caption{(Color online) The extrapolated superconducting correlations $\Phi_{xx}(r)$ and $\Phi_{yy}(r)$ in log-log scales on (A) $N=48\times 6$ and $N=24\time 8$ cylinders at $\delta=1/12$ and $J_2=0.5$, (B) $N=32\times 6$ and $N=24\times 8$ cylinders at $\delta=1/12$ and $J_2=0.5$ and (C) $N=48\times 6$ and $N=24\times 8$ at $\delta=1/12$ and $J_2=0.55$.} \label{FigA:SCa}
\end{figure}

Fig.\ref{FigA:SCa} shows more results including $\Phi_{xx}(r)$ and $\Phi_{yy}(r)$ on both $W=6$ and $W=8$ cylinders at different doping concentrations $\delta$ for $J_2=0.5$ and $J_2=0.55$. We find that in all cases $\Phi_{xx}(r)\approx \Phi_{yy}(r)$, and both are consistent with power-law decay $\Phi(r)\sim r^{-K_{\rm sc}}$ with similar exponents.\\

\section{PDW-like correlations in finite cylinders-some version}\label{SM:SC-PDW}%
Fig.\ref{Fig:SC} B show that at large separations, $r$, the pair-field correlators $\Phi_{\alpha\beta}({\bf r,r^\prime})=\langle\Delta^\dagger_\alpha({\bf r})\Delta_\beta({\bf r^\prime})\rangle$  for $W=4$ and 6  can be approximately described as the sum of two power law decaying components, a smoothly decaying piece, $\Phi_{\alpha\beta}^{(0)}({\bf r,r^\prime}) \sim |{\bf r-r^\prime}|^{-K_{\rm sc}}$, and an oscillatory piece, $\Phi_{\alpha\beta}^{({\bf Q})}({\bf r,r^\prime}) \sim \cos[{\bf Q\cdot (r-r^\prime)} +\theta]|{\bf r-r^\prime}|^{-K_{\rm sc}^\prime}$, with $K_{\rm sc}\approx K_{\rm sc}^\prime$. The oscillatory piece is absent, or at least much less prominent in the cylinder with $W=8$.

If such behavior were observed in a translationally invariant system - e.g. if it persisted to $L\to \infty$ , it would imply the existence of two distinct SC ordering tendencies, a uniform SC and a pair density wave (PDW).  However, as mentioned in the main text, we think this is not the correct interpretation of our observations. Various features of the data that are unnatural from this perspective include  1) The near equality of the two exponents would require fine-tuning.  2) The fact that $\bf Q$ is the same ordering vector that is seen in the CDW correlations would be reasonable in this scenario only if the CDW order were itself an induced order, in which case one would  expect that $K_{\rm cdw} = K_{\rm sc}+K_{\rm sc}^\prime$, an equality that is far from satisfied.  3) Moreover, we have checked for the case of $W=4$ that the magnitude of the oscillatory piece decreases with increasing $L$.

Instead, we have concluded that the oscillatory component of the SC correlations are a finite $L$ effect, reflecting the combined effects of a uniform SC and a CDW ordering tendency. The fundamental assumption is that the most relevant fields in this Luther-Emery liquid are (1) the uniform SC pair field $\psi_\alpha^{(0)}({\bf r})$ and (2) the CDW field $\rho_{\bf Q}(\bf r)$ where$$n({\bf r})-\bar{n}=a_0 \left(\rho_{\bf Q}({\bf r}) e^{i{\bf Q\cdot\bf r}}+\rho_{-\bf Q}({\bf r}) e^{-i{\bf Q\cdot\bf r}}\right).$$
Here both $\psi^{(0)}$ and $\rho_{\bf Q}$ are smooth varying fields, and $a_0$ is a non-universal constant.  Based on the operator product expansion, a modulating piece of the pair field is generated via 
$$\psi_\alpha^{(\bf Q)}(\bf r)=\rho_{\bf Q}({\bf r})\psi^{(0)}_\alpha({\bf r}) e^{i{\bf Q\cdot r}}+...$$
where ... denote the less relevant pieces. 
Thus the microscopic pair-field creation operator  
can be expanded as
\be
\Delta_\alpha({\bf r})&&=a_1~\psi^{(0)}_\alpha({\bf r})+a_2~\psi_\alpha^{(0)}({\bf r})\Big(\rho_{{\bf Q}}({\bf r}) e^{i{\bf Q\cdot r}} \nonumber\\&&+\rho_{-{\bf Q}}({\bf r}) e^{-i{\bf Q\cdot r}}\Big)+ \ldots \label{da}\ee
where  $a_1$ and $a_2$ are an non-universal amplitudes.  

Fig.\ref{Fig:CDW} A indicates that due to the "Friedel" oscillation induced by the boundary $\langle \rho_{\bf Q}({\bf r})\rangle $ and $\langle \rho_{\bf -Q}({\bf r})\rangle $ are non-zero hence we replace Eq.(\ref{da}) by 
\be
\Delta_\alpha({\bf r})&&=\Big[a_1+a_2 A({\bf r})\cos({\bf Q\cdot r}+\phi({\bf r}))\Big]\psi^{(0)}_\alpha({\bf r})\nonumber\\&&+ \ldots\label{da1}\ee
where $A(\bf r)$ is the amplitude and $\phi(\bf r)$ is the phase of the CDW, namely, 
\be \langle\rho_{{\bf Q}}({\bf r})\rangle e^{i{\bf Q\cdot r}} +\langle\rho_{-{\bf Q}}({\bf r})\rangle e^{-i{\bf Q\cdot r}}=A({\bf r})\cos({\bf Q\cdot r}+\phi(\bf r)).\nonumber\ee
Computing the pair-field correlation function using Eq.(\ref{da1}) leads to
\be
\Phi_{\alpha\beta}({\bf r,r^\prime})=\langle \psi_\alpha^{(0)*}({\bf r})\psi^{(0)}_\beta({\bf r^\prime})\rangle \Gamma({\bf r})\Gamma({\bf r^\prime})+\ldots,\label{da2}\ee 
where $$\Gamma({\bf r})=a_1+a_2A({\bf r})\cos({\bf Q\cdot r}+\phi({\bf r})),$$

and  $$\langle \psi_\alpha^{(0)*}({\bf r})\psi^{(0)}_\beta({\bf r^\prime})\rangle\sim |{\bf r-r^\prime}|^{-K_{\rm sc}}S_{\alpha\beta},$$with $S_{\alpha\beta}$ being the d-wave sign. 
In Fig.\ref{Fig:SC} 
 one of the point, say, $\bf r^\prime$ is fixed somewhere away from the boundary of the cylinder, say, ${\bf r_0}=(L/4,y_0)$ rendering  $\Gamma(\bf r^\prime)$ is a constant. The resulting pair field correlation function depends on $\Gamma({\bf r})$. We expect the boundary to induce a power-law decaying $A({\bf r})$ toward the interior of the cylinder, namely, $$A(x,y)=\bar{A}\left[\left({x-1\over L/4-1}\right)^{-K_c}+\left({L-x\over 3L/4}\right)^{-K_c}\right],$$where $\bar{A}$ is the value of $A({\bf r})$ at ${\bf r=r_0}$.
Consequently 
\be
&&\Phi_{\alpha\beta}({\bf r,r^\prime})\sim \nonumber\\&& S_{\alpha\beta}|{\bf r-r_0}|^{-K_{\rm sc}}\Big[a_1+a_2 A(x,y)\cos({\bf Q\cdot(r-r_0)}+\theta({\bf r}))\Big]\nonumber\\&&+\ldots.\nonumber\\
\label{da3}\ee
In Eq.(\ref{da3}) $\theta({\bf r})={\bf Q\cdot r_0}+\phi({\bf r}).$ 
Inspecting Fig.\ref{Fig:CDW}A  apparently neither 
$A(x,y)$ nor $\theta({\bf r})$ varies appreciably with ${\bf r}$ in the range of ${\bf r}$ plotted. If we replace these two quantities with constant, Eq. (\ref{da3}) implies a uniform SC component (the term proportional to $\alpha_1$) and an oscillatory SC component (the term proportional to $\alpha_2$). Importantly both components decay with the exponent $K_{\rm sc}$.   Note that the amplitude of the oscillatory component is proportional to $\bar{A}$ that vanishes as the finite size induced density oscillations vanish in the $L \to \infty$ limit. 

The higher order terms represented by $\ldots$ in Eq.(\ref{da3}) all decay with larger power-laws - including, for instance, an oscillatory term (which should  persist even in the $L\to\infty$ limit) with wave vector ${\bf Q}$ but which decays with power $K_{\rm sc}+K_{\rm cdw}$.  It is probably not feasible to extract such terms from presently achievable numerical data.

\begin{figure}
  \includegraphics[width=\linewidth]{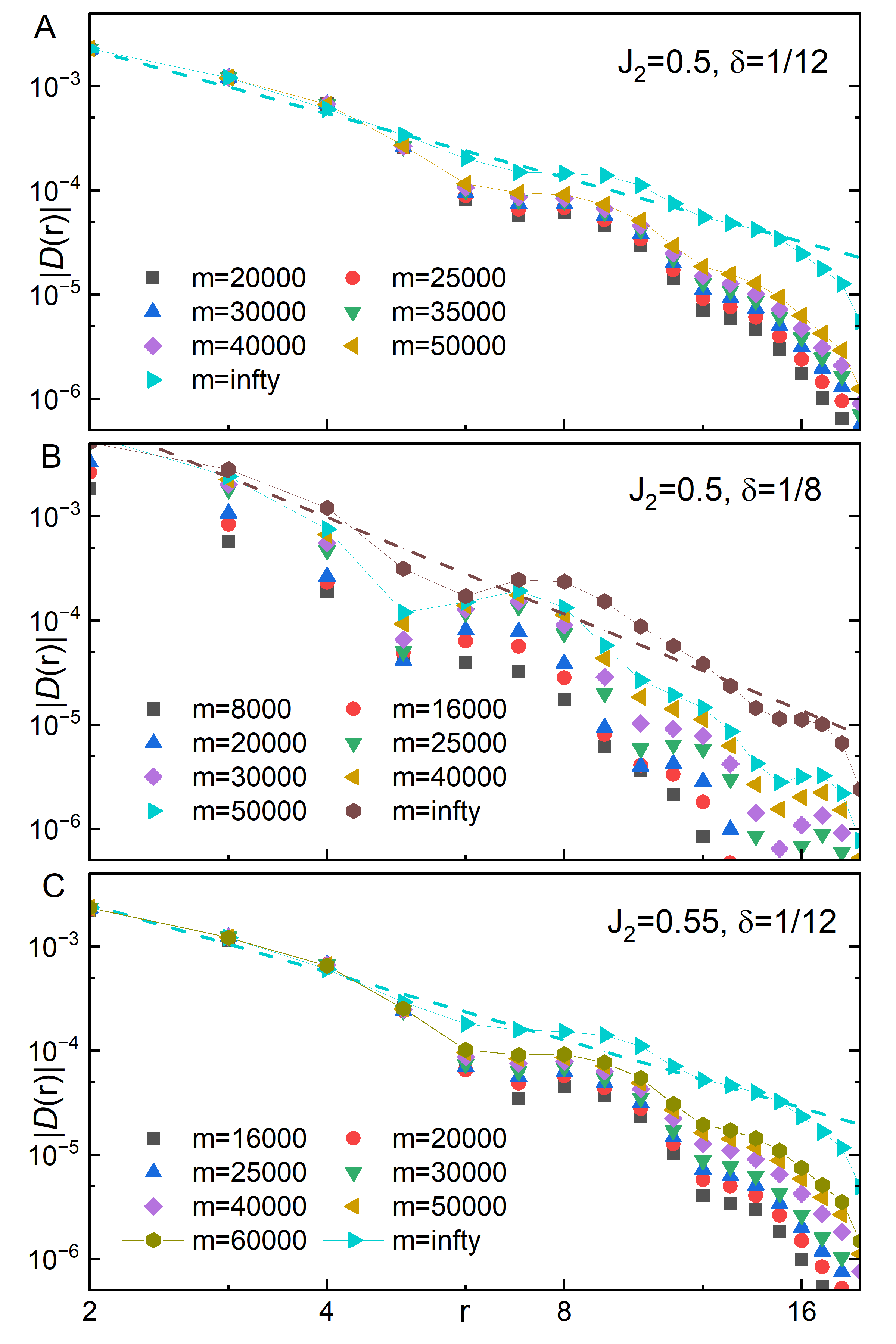}
\caption{(Color online) Log-log plots of $|D(r)|$ for $N=24\times 8$ cylinder at (A) $\delta=1/12$ with $J_2=0.5$, (B) $\delta=1/8$ with $J_2=0.5$, and (C) $\delta=1/12$ with $J_2=0.55$, where $r$ is the distance between two sites in the $\hat{x}$ direction. The dashed lines denote a power-law fit $D(r)\sim r^{-K_c}$. Here filled symbols represent the sign of $D(r)$ being $-$.}
\label{FigA:NN}
\end{figure}

\section{Site-charge density correlations}\label{SM:CDW}%
Fig.\ref{FigA:NN} shows the log-log plot of $|D(r)|$ (defined in Eq. \ref{Dofr})  versus $r$ for $N=24\times 8$ cylinders with $J_2=0.5$ at $\delta=1/12$, $J_2=0.5$ at $\delta=1/8$, and $J_2=0.55$ at $\delta=1/12$. Following similar procedure as $\Phi(r)$, the extrapolated $D(r)$ in the limit $m\rightarrow\infty$ is obtained using a second-order polynomial fit to the four data points associated with the largest $m$. As indicated by the dashed lines, the charge density-density correlations are consistent with a power-law decay $D(r)\propto r^{-K_{c}}$. To exclude the short-distance behavior and the boundary effects due to finite $L$, the data points with the smallest and the largest $r$ were omitted in obtaining this fit. For the $W=8$ cylinders with $J_2=0.5$, $\delta=1/12$, $\delta=1/8$ and $J_2=0.55$, $\delta=1/12$, the ``best fit'' values of $K_c$ all exceeds $2$. As discussed in the main text, under such conditions, the asymptotic behavior of $D(r)$ is expected to  be governed by the fluctuations of the acoustic modes. Following the same procedure, the ``best fit'' $K_c$ for $N=72\times 4$, $64\times 4$ and $48\times 6$ at various $J_2$ and $\delta$ are given in Table \ref{Table:Sum}.

In addition to $D(r)$, the exponent $K_c$ can also be extracted from the charge density oscillation $n(x)$ as shown in the main text.  The values of $K_c$ extracted this way for $N=72\times 4$, $N=64\times 4$ and $N=48\times 6$ cylinders for various $J_2$ and $\delta$ are also shown in Table \ref{Table:Sum}. Although the values of $K_c$ extracted from $D(r)$ is slightly different from that extracted from $n(x)$ (which may be caused by the boundary and finite-size effects), importantly they are all consistent with $K_c > K_{\rm sc}$.\\


\begin{thebibliography}{42}%
\makeatletter
\providecommand \@ifxundefined [1]{%
 \@ifx{#1\undefined}
}%
\providecommand \@ifnum [1]{%
 \ifnum #1\expandafter \@firstoftwo
 \else \expandafter \@secondoftwo
 \fi
}%
\providecommand \@ifx [1]{%
 \ifx #1\expandafter \@firstoftwo
 \else \expandafter \@secondoftwo
 \fi
}%
\providecommand \natexlab [1]{#1}%
\providecommand \enquote  [1]{``#1''}%
\providecommand \bibnamefont  [1]{#1}%
\providecommand \bibfnamefont [1]{#1}%
\providecommand \citenamefont [1]{#1}%
\providecommand \href@noop [0]{\@secondoftwo}%
\providecommand \href [0]{\begingroup \@sanitize@url \@href}%
\providecommand \@href[1]{\@@startlink{#1}\@@href}%
\providecommand \@@href[1]{\endgroup#1\@@endlink}%
\providecommand \@sanitize@url [0]{\catcode `\\12\catcode `\$12\catcode
  `\&12\catcode `\#12\catcode `\^12\catcode `\_12\catcode `\%12\relax}%
\providecommand \@@startlink[1]{}%
\providecommand \@@endlink[0]{}%
\providecommand \url  [0]{\begingroup\@sanitize@url \@url }%
\providecommand \@url [1]{\endgroup\@href {#1}{\urlprefix }}%
\providecommand \urlprefix  [0]{URL }%
\providecommand \Eprint [0]{\href }%
\providecommand \doibase [0]{http://dx.doi.org/}%
\providecommand \selectlanguage [0]{\@gobble}%
\providecommand \bibinfo  [0]{\@secondoftwo}%
\providecommand \bibfield  [0]{\@secondoftwo}%
\providecommand \translation [1]{[#1]}%
\providecommand \BibitemOpen [0]{}%
\providecommand \bibitemStop [0]{}%
\providecommand \bibitemNoStop [0]{.\EOS\space}%
\providecommand \EOS [0]{\spacefactor3000\relax}%
\providecommand \BibitemShut  [1]{\csname bibitem#1\endcsname}%
\let\auto@bib@innerbib\@empty
\bibitem [{\citenamefont {Anderson}(1987)}]{Anderson1987}%
  \BibitemOpen
  \bibfield  {author} {\bibinfo {author} {\bibfnamefont {P.~W.}\ \bibnamefont
  {Anderson}},\ }\href {\doibase 10.1126/science.235.4793.1196} {\bibfield
  {journal} {\bibinfo  {journal} {Science}\ }\textbf {\bibinfo {volume}
  {235}},\ \bibinfo {pages} {1196} (\bibinfo {year} {1987})}\BibitemShut
  {NoStop}%
\bibitem [{\citenamefont {Lee}\ \emph {et~al.}(2006)\citenamefont {Lee},
  \citenamefont {Nagaosa},\ and\ \citenamefont {Wen}}]{Lee2006}%
  \BibitemOpen
  \bibfield  {author} {\bibinfo {author} {\bibfnamefont {P.~A.}\ \bibnamefont
  {Lee}}, \bibinfo {author} {\bibfnamefont {N.}~\bibnamefont {Nagaosa}}, \ and\
  \bibinfo {author} {\bibfnamefont {X.-G.}\ \bibnamefont {Wen}},\ }\href
  {\doibase 10.1103/RevModPhys.78.17} {\bibfield  {journal} {\bibinfo
  {journal} {Rev. Mod. Phys.}\ }\textbf {\bibinfo {volume} {78}},\ \bibinfo
  {pages} {17} (\bibinfo {year} {2006})}\BibitemShut {NoStop}%
\bibitem [{\citenamefont {Weng}(2011)}]{Weng2011}%
  \BibitemOpen
  \bibfield  {author} {\bibinfo {author} {\bibfnamefont {Z.-Y.}\ \bibnamefont
  {Weng}},\ }\href {\doibase 10.1088/1367-2630/13/10/103039} {\bibfield
  {journal} {\bibinfo  {journal} {New Journal of Physics}\ }\textbf {\bibinfo
  {volume} {13}},\ \bibinfo {pages} {103039} (\bibinfo {year}
  {2011})}\BibitemShut {NoStop}%
\bibitem [{\citenamefont {{Broholm}}\ \emph {et~al.}(2019)\citenamefont
  {{Broholm}}, \citenamefont {{Cava}}, \citenamefont {{Kivelson}},
  \citenamefont {{Nocera}}, \citenamefont {{Norman}},\ and\ \citenamefont
  {{Senthil}}}]{broholmreview}%
  \BibitemOpen
  \bibfield  {author} {\bibinfo {author} {\bibfnamefont {C.}~\bibnamefont
  {{Broholm}}}, \bibinfo {author} {\bibfnamefont {R.~J.}\ \bibnamefont
  {{Cava}}}, \bibinfo {author} {\bibfnamefont {S.~A.}\ \bibnamefont
  {{Kivelson}}}, \bibinfo {author} {\bibfnamefont {D.~G.}\ \bibnamefont
  {{Nocera}}}, \bibinfo {author} {\bibfnamefont {M.~R.}\ \bibnamefont
  {{Norman}}}, \ and\ \bibinfo {author} {\bibfnamefont {T.}~\bibnamefont
  {{Senthil}}},\ }\href@noop {} {\bibfield  {journal} {\bibinfo  {journal}
  {arXiv e-prints}\ ,\ \bibinfo {eid} {arXiv:1905.07040}} (\bibinfo {year}
  {2019})},\ \Eprint {http://arxiv.org/abs/1905.07040} {arXiv:1905.07040
  [cond-mat.str-el]} \BibitemShut {NoStop}%
\bibitem [{\citenamefont {Arovas}\ \emph {et~al.}(2022)\citenamefont {Arovas},
  \citenamefont {Berg}, \citenamefont {Kivelson},\ and\ \citenamefont
  {Raghu}}]{Arovas2022}%
  \BibitemOpen
  \bibfield  {author} {\bibinfo {author} {\bibfnamefont {D.~P.}\ \bibnamefont
  {Arovas}}, \bibinfo {author} {\bibfnamefont {E.}~\bibnamefont {Berg}},
  \bibinfo {author} {\bibfnamefont {S.~A.}\ \bibnamefont {Kivelson}}, \ and\
  \bibinfo {author} {\bibfnamefont {S.}~\bibnamefont {Raghu}},\ }\href
  {\doibase 10.1146/annurev-conmatphys-031620-102024} {\bibfield  {journal}
  {\bibinfo  {journal} {Annual Review of Condensed Matter Physics}\ }\textbf
  {\bibinfo {volume} {13}},\ \bibinfo {pages} {239} (\bibinfo {year}
  {2022})}\BibitemShut {NoStop}%
\bibitem [{\citenamefont {Weng}\ \emph {et~al.}(1999)\citenamefont {Weng},
  \citenamefont {Sheng},\ and\ \citenamefont {Ting}}]{Weng1999}%
  \BibitemOpen
  \bibfield  {author} {\bibinfo {author} {\bibfnamefont {Z.~Y.}\ \bibnamefont
  {Weng}}, \bibinfo {author} {\bibfnamefont {D.~N.}\ \bibnamefont {Sheng}}, \
  and\ \bibinfo {author} {\bibfnamefont {C.~S.}\ \bibnamefont {Ting}},\ }\href
  {\doibase 10.1103/PhysRevB.59.8943} {\bibfield  {journal} {\bibinfo
  {journal} {Phys. Rev. B}\ }\textbf {\bibinfo {volume} {59}},\ \bibinfo
  {pages} {8943} (\bibinfo {year} {1999})}\BibitemShut {NoStop}%
\bibitem [{\citenamefont {Qin}\ \emph {et~al.}(2022)\citenamefont {Qin},
  \citenamefont {Schäfer}, \citenamefont {Andergassen}, \citenamefont
  {Corboz},\ and\ \citenamefont {Gull}}]{Qin2022}%
  \BibitemOpen
  \bibfield  {author} {\bibinfo {author} {\bibfnamefont {M.}~\bibnamefont
  {Qin}}, \bibinfo {author} {\bibfnamefont {T.}~\bibnamefont {Schäfer}},
  \bibinfo {author} {\bibfnamefont {S.}~\bibnamefont {Andergassen}}, \bibinfo
  {author} {\bibfnamefont {P.}~\bibnamefont {Corboz}}, \ and\ \bibinfo {author}
  {\bibfnamefont {E.}~\bibnamefont {Gull}},\ }\href {\doibase
  10.1146/annurev-conmatphys-090921-033948} {\bibfield  {journal} {\bibinfo
  {journal} {Annual Review of Condensed Matter Physics}\ }\textbf {\bibinfo
  {volume} {13}},\ \bibinfo {pages} {275} (\bibinfo {year} {2022})}\BibitemShut
  {NoStop}%
\bibitem [{\citenamefont {Qin}\ \emph {et~al.}(2020)\citenamefont {Qin},
  \citenamefont {Chung}, \citenamefont {Shi}, \citenamefont {Vitali},
  \citenamefont {Hubig}, \citenamefont {Schollw\"ock}, \citenamefont {White},\
  and\ \citenamefont {Zhang}}]{AbsenceSC}%
  \BibitemOpen
  \bibfield  {author} {\bibinfo {author} {\bibfnamefont {M.}~\bibnamefont
  {Qin}}, \bibinfo {author} {\bibfnamefont {C.-M.}\ \bibnamefont {Chung}},
  \bibinfo {author} {\bibfnamefont {H.}~\bibnamefont {Shi}}, \bibinfo {author}
  {\bibfnamefont {E.}~\bibnamefont {Vitali}}, \bibinfo {author} {\bibfnamefont
  {C.}~\bibnamefont {Hubig}}, \bibinfo {author} {\bibfnamefont
  {U.}~\bibnamefont {Schollw\"ock}}, \bibinfo {author} {\bibfnamefont {S.~R.}\
  \bibnamefont {White}}, \ and\ \bibinfo {author} {\bibfnamefont
  {S.}~\bibnamefont {Zhang}} (\bibinfo {collaboration} {Simons Collaboration on
  the Many-Electron Problem}),\ }\href {\doibase 10.1103/PhysRevX.10.031016}
  {\bibfield  {journal} {\bibinfo  {journal} {Phys. Rev. X}\ }\textbf {\bibinfo
  {volume} {10}},\ \bibinfo {pages} {031016} (\bibinfo {year}
  {2020})}\BibitemShut {NoStop}%
\bibitem [{\citenamefont {White}\ and\ \citenamefont
  {Scalapino}(1999)}]{White1999}%
  \BibitemOpen
  \bibfield  {author} {\bibinfo {author} {\bibfnamefont {S.~R.}\ \bibnamefont
  {White}}\ and\ \bibinfo {author} {\bibfnamefont {D.~J.}\ \bibnamefont
  {Scalapino}},\ }\href {\doibase 10.1103/PhysRevB.60.R753} {\bibfield
  {journal} {\bibinfo  {journal} {Phys. Rev. B}\ }\textbf {\bibinfo {volume}
  {60}},\ \bibinfo {pages} {R753} (\bibinfo {year} {1999})}\BibitemShut
  {NoStop}%
\bibitem [{\citenamefont {Scalapino}\ and\ \citenamefont
  {White}(2012)}]{scalapinowhitereview}%
  \BibitemOpen
  \bibfield  {author} {\bibinfo {author} {\bibfnamefont {D.}~\bibnamefont
  {Scalapino}}\ and\ \bibinfo {author} {\bibfnamefont {S.}~\bibnamefont
  {White}},\ }\href {\doibase https://doi.org/10.1016/j.physc.2012.04.004}
  {\bibfield  {journal} {\bibinfo  {journal} {Physica C: Superconductivity}\
  }\textbf {\bibinfo {volume} {481}},\ \bibinfo {pages} {146 } (\bibinfo {year}
  {2012})},\ \bibinfo {note} {stripes and Electronic Liquid Crystals in
  Strongly Correlated Materials}\BibitemShut {NoStop}%
\bibitem [{\citenamefont {Dodaro}\ \emph {et~al.}(2017)\citenamefont {Dodaro},
  \citenamefont {Jiang},\ and\ \citenamefont {Kivelson}}]{Dodaro2017}%
  \BibitemOpen
  \bibfield  {author} {\bibinfo {author} {\bibfnamefont {J.~F.}\ \bibnamefont
  {Dodaro}}, \bibinfo {author} {\bibfnamefont {H.-C.}\ \bibnamefont {Jiang}}, \
  and\ \bibinfo {author} {\bibfnamefont {S.~A.}\ \bibnamefont {Kivelson}},\
  }\href {\doibase 10.1103/PhysRevB.95.155116} {\bibfield  {journal} {\bibinfo
  {journal} {Phys. Rev. B}\ }\textbf {\bibinfo {volume} {95}},\ \bibinfo
  {pages} {155116} (\bibinfo {year} {2017})}\BibitemShut {NoStop}%
\bibitem [{Note1()}]{Note1}%
  \BibitemOpen
  \bibinfo {note} {Similar conclusions concerning competing orders quenching SC
  were reached on the basis of variational auxiliary field quantum Monte Carlo
  calculations in Ref. \cite {sorella,ZhangStripes,Xiao2023}.}\BibitemShut
  {Stop}%
\bibitem [{\citenamefont {Zheng}\ \emph {et~al.}(2017)\citenamefont {Zheng},
  \citenamefont {Chung}, \citenamefont {Corboz}, \citenamefont {Ehlers},
  \citenamefont {Qin}, \citenamefont {Noack}, \citenamefont {Shi},
  \citenamefont {White}, \citenamefont {Zhang},\ and\ \citenamefont
  {Chan}}]{Zheng2017}%
  \BibitemOpen
  \bibfield  {author} {\bibinfo {author} {\bibfnamefont {B.-X.}\ \bibnamefont
  {Zheng}}, \bibinfo {author} {\bibfnamefont {C.-M.}\ \bibnamefont {Chung}},
  \bibinfo {author} {\bibfnamefont {P.}~\bibnamefont {Corboz}}, \bibinfo
  {author} {\bibfnamefont {G.}~\bibnamefont {Ehlers}}, \bibinfo {author}
  {\bibfnamefont {M.-P.}\ \bibnamefont {Qin}}, \bibinfo {author} {\bibfnamefont
  {R.~M.}\ \bibnamefont {Noack}}, \bibinfo {author} {\bibfnamefont
  {H.}~\bibnamefont {Shi}}, \bibinfo {author} {\bibfnamefont {S.~R.}\
  \bibnamefont {White}}, \bibinfo {author} {\bibfnamefont {S.}~\bibnamefont
  {Zhang}}, \ and\ \bibinfo {author} {\bibfnamefont {G.~K.-L.}\ \bibnamefont
  {Chan}},\ }\href@noop {} {\bibfield  {journal} {\bibinfo  {journal}
  {Science}\ }\textbf {\bibinfo {volume} {358}},\ \bibinfo {pages} {1155}
  (\bibinfo {year} {2017})}\BibitemShut {NoStop}%
\bibitem [{\citenamefont {Jiang}\ \emph
  {et~al.}(2020{\natexlab{a}})\citenamefont {Jiang}, \citenamefont {Zaanen},
  \citenamefont {Devereaux},\ and\ \citenamefont {Jiang}}]{Jiang2020prr}%
  \BibitemOpen
  \bibfield  {author} {\bibinfo {author} {\bibfnamefont {Y.-F.}\ \bibnamefont
  {Jiang}}, \bibinfo {author} {\bibfnamefont {J.}~\bibnamefont {Zaanen}},
  \bibinfo {author} {\bibfnamefont {T.~P.}\ \bibnamefont {Devereaux}}, \ and\
  \bibinfo {author} {\bibfnamefont {H.-C.}\ \bibnamefont {Jiang}},\ }\href
  {\doibase 10.1103/PhysRevResearch.2.033073} {\bibfield  {journal} {\bibinfo
  {journal} {Phys. Rev. Research}\ }\textbf {\bibinfo {volume} {2}},\ \bibinfo
  {pages} {033073} (\bibinfo {year} {2020}{\natexlab{a}})}\BibitemShut
  {NoStop}%
\bibitem [{\citenamefont {Gong}\ \emph {et~al.}(2021)\citenamefont {Gong},
  \citenamefont {Zhu},\ and\ \citenamefont {Sheng}}]{Gong2021}%
  \BibitemOpen
  \bibfield  {author} {\bibinfo {author} {\bibfnamefont {S.}~\bibnamefont
  {Gong}}, \bibinfo {author} {\bibfnamefont {W.}~\bibnamefont {Zhu}}, \ and\
  \bibinfo {author} {\bibfnamefont {D.~N.}\ \bibnamefont {Sheng}},\ }\href
  {\doibase 10.1103/PhysRevLett.127.097003} {\bibfield  {journal} {\bibinfo
  {journal} {Phys. Rev. Lett.}\ }\textbf {\bibinfo {volume} {127}},\ \bibinfo
  {pages} {097003} (\bibinfo {year} {2021})}\BibitemShut {NoStop}%
\bibitem [{\citenamefont {Jiang}\ \emph {et~al.}(2021)\citenamefont {Jiang},
  \citenamefont {Scalapino},\ and\ \citenamefont {White}}]{Jiang2021White}%
  \BibitemOpen
  \bibfield  {author} {\bibinfo {author} {\bibfnamefont {S.}~\bibnamefont
  {Jiang}}, \bibinfo {author} {\bibfnamefont {D.~J.}\ \bibnamefont
  {Scalapino}}, \ and\ \bibinfo {author} {\bibfnamefont {S.~R.}\ \bibnamefont
  {White}},\ }\href@noop {} {\bibfield  {journal} {\bibinfo  {journal} {Proc.
  Natl. Acad. Sci. U.S.A.}\ }\textbf {\bibinfo {volume} {118}},\ \bibinfo
  {pages} {e2109978118} (\bibinfo {year} {2021})}\BibitemShut {NoStop}%
\bibitem [{\citenamefont {Jiang}\ \emph {et~al.}(2018)\citenamefont {Jiang},
  \citenamefont {Weng},\ and\ \citenamefont {Kivelson}}]{Jiang2018tJ}%
  \BibitemOpen
  \bibfield  {author} {\bibinfo {author} {\bibfnamefont {H.-C.}\ \bibnamefont
  {Jiang}}, \bibinfo {author} {\bibfnamefont {Z.-Y.}\ \bibnamefont {Weng}}, \
  and\ \bibinfo {author} {\bibfnamefont {S.~A.}\ \bibnamefont {Kivelson}},\
  }\href@noop {} {\bibfield  {journal} {\bibinfo  {journal} {Phys. Rev. B}\
  }\textbf {\bibinfo {volume} {98}},\ \bibinfo {pages} {140505} (\bibinfo
  {year} {2018})}\BibitemShut {NoStop}%
\bibitem [{\citenamefont {Jiang}\ \emph
  {et~al.}(2020{\natexlab{b}})\citenamefont {Jiang}, \citenamefont {Chen},\
  and\ \citenamefont {Weng}}]{Jiang2020tJ}%
  \BibitemOpen
  \bibfield  {author} {\bibinfo {author} {\bibfnamefont {H.-C.}\ \bibnamefont
  {Jiang}}, \bibinfo {author} {\bibfnamefont {S.}~\bibnamefont {Chen}}, \ and\
  \bibinfo {author} {\bibfnamefont {Z.-Y.}\ \bibnamefont {Weng}},\ }\href
  {\doibase 10.1103/PhysRevB.102.104512} {\bibfield  {journal} {\bibinfo
  {journal} {Phys. Rev. B}\ }\textbf {\bibinfo {volume} {102}},\ \bibinfo
  {pages} {104512} (\bibinfo {year} {2020}{\natexlab{b}})}\BibitemShut
  {NoStop}%
\bibitem [{\citenamefont {Chung}\ \emph {et~al.}(2020)\citenamefont {Chung},
  \citenamefont {Qin}, \citenamefont {Zhang}, \citenamefont {Schollw\"ock},\
  and\ \citenamefont {White}}]{Chung2020}%
  \BibitemOpen
  \bibfield  {author} {\bibinfo {author} {\bibfnamefont {C.-M.}\ \bibnamefont
  {Chung}}, \bibinfo {author} {\bibfnamefont {M.}~\bibnamefont {Qin}}, \bibinfo
  {author} {\bibfnamefont {S.}~\bibnamefont {Zhang}}, \bibinfo {author}
  {\bibfnamefont {U.}~\bibnamefont {Schollw\"ock}}, \ and\ \bibinfo {author}
  {\bibfnamefont {S.~R.}\ \bibnamefont {White}} (\bibinfo {collaboration} {The
  Simons Collaboration on the Many-Electron Problem}),\ }\href {\doibase
  10.1103/PhysRevB.102.041106} {\bibfield  {journal} {\bibinfo  {journal}
  {Phys. Rev. B}\ }\textbf {\bibinfo {volume} {102}},\ \bibinfo {pages}
  {041106} (\bibinfo {year} {2020})}\BibitemShut {NoStop}%
\bibitem [{\citenamefont {Jiang}\ and\ \citenamefont
  {Devereaux}(2019)}]{Jiang2019Hub}%
  \BibitemOpen
  \bibfield  {author} {\bibinfo {author} {\bibfnamefont {H.-C.}\ \bibnamefont
  {Jiang}}\ and\ \bibinfo {author} {\bibfnamefont {T.~P.}\ \bibnamefont
  {Devereaux}},\ }\href {\doibase 10.1126/science.aal5304} {\bibfield
  {journal} {\bibinfo  {journal} {Science}\ }\textbf {\bibinfo {volume}
  {365}},\ \bibinfo {pages} {1424} (\bibinfo {year} {2019})}\BibitemShut
  {NoStop}%
\bibitem [{\citenamefont {Jiang}\ and\ \citenamefont
  {Kivelson}(2021)}]{Jiang2021}%
  \BibitemOpen
  \bibfield  {author} {\bibinfo {author} {\bibfnamefont {H.-C.}\ \bibnamefont
  {Jiang}}\ and\ \bibinfo {author} {\bibfnamefont {S.~A.}\ \bibnamefont
  {Kivelson}},\ }\href {\doibase 10.1103/PhysRevLett.127.097002} {\bibfield
  {journal} {\bibinfo  {journal} {Phys. Rev. Lett.}\ }\textbf {\bibinfo
  {volume} {127}},\ \bibinfo {pages} {097002} (\bibinfo {year}
  {2021})}\BibitemShut {NoStop}%
\bibitem [{\citenamefont {Peng}\ \emph {et~al.}(2022)\citenamefont {Peng},
  \citenamefont {Wang}, \citenamefont {Wen}, \citenamefont {Lee}, \citenamefont
  {Devereaux},\ and\ \citenamefont {Jiang}}]{Peng2022}%
  \BibitemOpen
  \bibfield  {author} {\bibinfo {author} {\bibfnamefont {C.}~\bibnamefont
  {Peng}}, \bibinfo {author} {\bibfnamefont {Y.}~\bibnamefont {Wang}}, \bibinfo
  {author} {\bibfnamefont {J.}~\bibnamefont {Wen}}, \bibinfo {author}
  {\bibfnamefont {Y.}~\bibnamefont {Lee}}, \bibinfo {author} {\bibfnamefont
  {T.}~\bibnamefont {Devereaux}}, \ and\ \bibinfo {author} {\bibfnamefont
  {H.-C.}\ \bibnamefont {Jiang}},\ }\href {\doibase 10.48550/ARXIV.2206.03486}
  {\enquote {\bibinfo {title} {Enhanced superconductivity by near-neighbor
  attraction in the doped hubbard model},}\ } (\bibinfo {year}
  {2022})\BibitemShut {NoStop}%
\bibitem [{\citenamefont {Damascelli}\ \emph {et~al.}(2003)\citenamefont
  {Damascelli}, \citenamefont {Hussain},\ and\ \citenamefont
  {Shen}}]{damascelli}%
  \BibitemOpen
  \bibfield  {author} {\bibinfo {author} {\bibfnamefont {A.}~\bibnamefont
  {Damascelli}}, \bibinfo {author} {\bibfnamefont {Z.}~\bibnamefont {Hussain}},
  \ and\ \bibinfo {author} {\bibfnamefont {Z.-X.}\ \bibnamefont {Shen}},\
  }\href {\doibase 10.1103/RevModPhys.75.473} {\bibfield  {journal} {\bibinfo
  {journal} {Rev. Mod. Phys.}\ }\textbf {\bibinfo {volume} {75}},\ \bibinfo
  {pages} {473} (\bibinfo {year} {2003})}\BibitemShut {NoStop}%
\bibitem [{\citenamefont {Gong}\ \emph {et~al.}(2014)\citenamefont {Gong},
  \citenamefont {Zhu}, \citenamefont {Sheng}, \citenamefont {Motrunich},\ and\
  \citenamefont {Fisher}}]{Gong2014}%
  \BibitemOpen
  \bibfield  {author} {\bibinfo {author} {\bibfnamefont {S.-S.}\ \bibnamefont
  {Gong}}, \bibinfo {author} {\bibfnamefont {W.}~\bibnamefont {Zhu}}, \bibinfo
  {author} {\bibfnamefont {D.~N.}\ \bibnamefont {Sheng}}, \bibinfo {author}
  {\bibfnamefont {O.~I.}\ \bibnamefont {Motrunich}}, \ and\ \bibinfo {author}
  {\bibfnamefont {M.~P.~A.}\ \bibnamefont {Fisher}},\ }\href {\doibase
  10.1103/PhysRevLett.113.027201} {\bibfield  {journal} {\bibinfo  {journal}
  {Phys. Rev. Lett.}\ }\textbf {\bibinfo {volume} {113}},\ \bibinfo {pages}
  {027201} (\bibinfo {year} {2014})}\BibitemShut {NoStop}%
\bibitem [{\citenamefont {Wang}\ and\ \citenamefont
  {Sandvik}(2018)}]{Wang2018}%
  \BibitemOpen
  \bibfield  {author} {\bibinfo {author} {\bibfnamefont {L.}~\bibnamefont
  {Wang}}\ and\ \bibinfo {author} {\bibfnamefont {A.~W.}\ \bibnamefont
  {Sandvik}},\ }\href {\doibase 10.1103/PhysRevLett.121.107202} {\bibfield
  {journal} {\bibinfo  {journal} {Phys. Rev. Lett.}\ }\textbf {\bibinfo
  {volume} {121}},\ \bibinfo {pages} {107202} (\bibinfo {year}
  {2018})}\BibitemShut {NoStop}%
\bibitem [{\citenamefont {Liu}\ \emph {et~al.}(2022{\natexlab{a}})\citenamefont
  {Liu}, \citenamefont {Gong}, \citenamefont {Li}, \citenamefont {Poilblanc},
  \citenamefont {Chen},\ and\ \citenamefont {Gu}}]{Liu2022SciBul}%
  \BibitemOpen
  \bibfield  {author} {\bibinfo {author} {\bibfnamefont {W.-Y.}\ \bibnamefont
  {Liu}}, \bibinfo {author} {\bibfnamefont {S.-S.}\ \bibnamefont {Gong}},
  \bibinfo {author} {\bibfnamefont {Y.-B.}\ \bibnamefont {Li}}, \bibinfo
  {author} {\bibfnamefont {D.}~\bibnamefont {Poilblanc}}, \bibinfo {author}
  {\bibfnamefont {W.-Q.}\ \bibnamefont {Chen}}, \ and\ \bibinfo {author}
  {\bibfnamefont {Z.-C.}\ \bibnamefont {Gu}},\ }\href {\doibase
  https://doi.org/10.1016/j.scib.2022.03.010} {\bibfield  {journal} {\bibinfo
  {journal} {Science Bulletin}\ }\textbf {\bibinfo {volume} {67}},\ \bibinfo
  {pages} {1034} (\bibinfo {year} {2022}{\natexlab{a}})}\BibitemShut {NoStop}%
\bibitem [{\citenamefont {Liu}\ \emph {et~al.}(2022{\natexlab{b}})\citenamefont
  {Liu}, \citenamefont {Hasik}, \citenamefont {Gong}, \citenamefont
  {Poilblanc}, \citenamefont {Chen},\ and\ \citenamefont {Gu}}]{Liu2022PRX}%
  \BibitemOpen
  \bibfield  {author} {\bibinfo {author} {\bibfnamefont {W.-Y.}\ \bibnamefont
  {Liu}}, \bibinfo {author} {\bibfnamefont {J.}~\bibnamefont {Hasik}}, \bibinfo
  {author} {\bibfnamefont {S.-S.}\ \bibnamefont {Gong}}, \bibinfo {author}
  {\bibfnamefont {D.}~\bibnamefont {Poilblanc}}, \bibinfo {author}
  {\bibfnamefont {W.-Q.}\ \bibnamefont {Chen}}, \ and\ \bibinfo {author}
  {\bibfnamefont {Z.-C.}\ \bibnamefont {Gu}},\ }\href {\doibase
  10.1103/PhysRevX.12.031039} {\bibfield  {journal} {\bibinfo  {journal} {Phys.
  Rev. X}\ }\textbf {\bibinfo {volume} {12}},\ \bibinfo {pages} {031039}
  (\bibinfo {year} {2022}{\natexlab{b}})}\BibitemShut {NoStop}%
\bibitem [{\citenamefont {Jiang}\ \emph {et~al.}(2012)\citenamefont {Jiang},
  \citenamefont {Yao},\ and\ \citenamefont {Balents}}]{Jiang2012}%
  \BibitemOpen
  \bibfield  {author} {\bibinfo {author} {\bibfnamefont {H.-C.}\ \bibnamefont
  {Jiang}}, \bibinfo {author} {\bibfnamefont {H.}~\bibnamefont {Yao}}, \ and\
  \bibinfo {author} {\bibfnamefont {L.}~\bibnamefont {Balents}},\ }\href
  {\doibase 10.1103/PhysRevB.86.024424} {\bibfield  {journal} {\bibinfo
  {journal} {Phys. Rev. B}\ }\textbf {\bibinfo {volume} {86}},\ \bibinfo
  {pages} {024424} (\bibinfo {year} {2012})}\BibitemShut {NoStop}%
\bibitem [{\citenamefont {{Gannot}}\ and\ \citenamefont
  {{Kivelson}}(2022)}]{gannot}%
  \BibitemOpen
  \bibfield  {author} {\bibinfo {author} {\bibfnamefont {Y.}~\bibnamefont
  {{Gannot}}}\ and\ \bibinfo {author} {\bibfnamefont {S.~A.}\ \bibnamefont
  {{Kivelson}}},\ }\href@noop {} {\bibfield  {journal} {\bibinfo  {journal}
  {arXiv e-prints}\ ,\ \bibinfo {eid} {arXiv:2206.13519}} (\bibinfo {year}
  {2022})},\ \Eprint {http://arxiv.org/abs/2206.13519} {arXiv:2206.13519
  [cond-mat.str-el]} \BibitemShut {NoStop}%
\bibitem [{\citenamefont {Rokhsar}\ and\ \citenamefont
  {Kivelson}(1988)}]{rokhsarandme}%
  \BibitemOpen
  \bibfield  {author} {\bibinfo {author} {\bibfnamefont {D.~S.}\ \bibnamefont
  {Rokhsar}}\ and\ \bibinfo {author} {\bibfnamefont {S.~A.}\ \bibnamefont
  {Kivelson}},\ }\href {\doibase 10.1103/PhysRevLett.61.2376} {\bibfield
  {journal} {\bibinfo  {journal} {Phys. Rev. Lett.}\ }\textbf {\bibinfo
  {volume} {61}},\ \bibinfo {pages} {2376} (\bibinfo {year}
  {1988})}\BibitemShut {NoStop}%
\bibitem [{\citenamefont {Jiang}\ and\ \citenamefont
  {Kivelson}(2022)}]{Jiang2022pnas}%
  \BibitemOpen
  \bibfield  {author} {\bibinfo {author} {\bibfnamefont {H.-C.}\ \bibnamefont
  {Jiang}}\ and\ \bibinfo {author} {\bibfnamefont {S.~A.}\ \bibnamefont
  {Kivelson}},\ }\href {\doibase 10.1073/pnas.2109406119} {\bibfield  {journal}
  {\bibinfo  {journal} {Proceedings of the National Academy of Sciences}\
  }\textbf {\bibinfo {volume} {119}},\ \bibinfo {pages} {e2109406119} (\bibinfo
  {year} {2022})}\BibitemShut {NoStop}%
\bibitem [{\citenamefont {Li}\ \emph {et~al.}(2023)\citenamefont {Li},
  \citenamefont {Louie},\ and\ \citenamefont {Lee}}]{Li2023}%
  \BibitemOpen
  \bibfield  {author} {\bibinfo {author} {\bibfnamefont {Z.-X.}\ \bibnamefont
  {Li}}, \bibinfo {author} {\bibfnamefont {S.~G.}\ \bibnamefont {Louie}}, \
  and\ \bibinfo {author} {\bibfnamefont {D.-H.}\ \bibnamefont {Lee}},\ }\href
  {\doibase 10.1103/PhysRevB.107.L041103} {\bibfield  {journal} {\bibinfo
  {journal} {Phys. Rev. B}\ }\textbf {\bibinfo {volume} {107}},\ \bibinfo
  {pages} {L041103} (\bibinfo {year} {2023})}\BibitemShut {NoStop}%
\bibitem [{\citenamefont {White}(1992)}]{White1992}%
  \BibitemOpen
  \bibfield  {author} {\bibinfo {author} {\bibfnamefont {S.~R.}\ \bibnamefont
  {White}},\ }\href@noop {} {\bibfield  {journal} {\bibinfo  {journal} {Phys.
  Rev. Lett.}\ }\textbf {\bibinfo {volume} {69}},\ \bibinfo {pages} {2863}
  (\bibinfo {year} {1992})}\BibitemShut {NoStop}%
\bibitem [{\citenamefont {White}\ \emph {et~al.}(2002)\citenamefont {White},
  \citenamefont {Affleck},\ and\ \citenamefont {Scalapino}}]{White2002}%
  \BibitemOpen
  \bibfield  {author} {\bibinfo {author} {\bibfnamefont {S.~R.}\ \bibnamefont
  {White}}, \bibinfo {author} {\bibfnamefont {I.}~\bibnamefont {Affleck}}, \
  and\ \bibinfo {author} {\bibfnamefont {D.~J.}\ \bibnamefont {Scalapino}},\
  }\href {\doibase 10.1103/PhysRevB.65.165122} {\bibfield  {journal} {\bibinfo
  {journal} {Phys. Rev. B}\ }\textbf {\bibinfo {volume} {65}},\ \bibinfo
  {pages} {165122} (\bibinfo {year} {2002})}\BibitemShut {NoStop}%
\bibitem [{\citenamefont {{Tasaki}}(1989)}]{tasaki}%
  \BibitemOpen
  \bibfield  {author} {\bibinfo {author} {\bibfnamefont {H.}~\bibnamefont
  {{Tasaki}}},\ }\href {\doibase 10.1103/PhysRevB.40.9192} {\bibfield
  {journal} {\bibinfo  {journal} {\prb}\ }\textbf {\bibinfo {volume} {40}},\
  \bibinfo {pages} {9192} (\bibinfo {year} {1989})}\BibitemShut {NoStop}%
\bibitem [{\citenamefont {Kane}\ \emph {et~al.}(1989)\citenamefont {Kane},
  \citenamefont {Lee},\ and\ \citenamefont {Read}}]{kanelee}%
  \BibitemOpen
  \bibfield  {author} {\bibinfo {author} {\bibfnamefont {C.~L.}\ \bibnamefont
  {Kane}}, \bibinfo {author} {\bibfnamefont {P.~A.}\ \bibnamefont {Lee}}, \
  and\ \bibinfo {author} {\bibfnamefont {N.}~\bibnamefont {Read}},\ }\href
  {\doibase 10.1103/PhysRevB.39.6880} {\bibfield  {journal} {\bibinfo
  {journal} {Phys. Rev. B}\ }\textbf {\bibinfo {volume} {39}},\ \bibinfo
  {pages} {6880} (\bibinfo {year} {1989})}\BibitemShut {NoStop}%
\bibitem [{\citenamefont {{Kim}}(2022)}]{kyungsu}%
  \BibitemOpen
  \bibfield  {author} {\bibinfo {author} {\bibfnamefont {K.-S.}\ \bibnamefont
  {{Kim}}},\ }\href@noop {} {\bibfield  {journal} {\bibinfo  {journal} {arXiv
  e-prints}\ ,\ \bibinfo {eid} {arXiv:2207.09498}} (\bibinfo {year} {2022})},\
  \Eprint {http://arxiv.org/abs/2207.09498} {arXiv:2207.09498
  [cond-mat.str-el]} \BibitemShut {NoStop}%
\bibitem [{\citenamefont {Martins}\ \emph {et~al.}(2001)\citenamefont
  {Martins}, \citenamefont {Xavier}, \citenamefont {Arrachea},\ and\
  \citenamefont {Dagotto}}]{Martins2001}%
  \BibitemOpen
  \bibfield  {author} {\bibinfo {author} {\bibfnamefont {G.~B.}\ \bibnamefont
  {Martins}}, \bibinfo {author} {\bibfnamefont {J.~C.}\ \bibnamefont {Xavier}},
  \bibinfo {author} {\bibfnamefont {L.}~\bibnamefont {Arrachea}}, \ and\
  \bibinfo {author} {\bibfnamefont {E.}~\bibnamefont {Dagotto}},\ }\href
  {\doibase 10.1103/PhysRevB.64.180513} {\bibfield  {journal} {\bibinfo
  {journal} {Phys. Rev. B}\ }\textbf {\bibinfo {volume} {64}},\ \bibinfo
  {pages} {180513} (\bibinfo {year} {2001})}\BibitemShut {NoStop}%
\bibitem [{\citenamefont {Ioffe}\ and\ \citenamefont
  {Larkin}(1989)}]{IoffeLarkin}%
  \BibitemOpen
  \bibfield  {author} {\bibinfo {author} {\bibfnamefont {L.~B.}\ \bibnamefont
  {Ioffe}}\ and\ \bibinfo {author} {\bibfnamefont {A.~I.}\ \bibnamefont
  {Larkin}},\ }\href {\doibase 10.1103/PhysRevB.40.6941} {\bibfield  {journal}
  {\bibinfo  {journal} {Phys. Rev. B}\ }\textbf {\bibinfo {volume} {40}},\
  \bibinfo {pages} {6941} (\bibinfo {year} {1989})}\BibitemShut {NoStop}%
\bibitem [{\citenamefont {{Sorella}}(2021)}]{sorella}%
  \BibitemOpen
  \bibfield  {author} {\bibinfo {author} {\bibfnamefont {S.}~\bibnamefont
  {{Sorella}}},\ }\href@noop {} {\bibfield  {journal} {\bibinfo  {journal}
  {arXiv e-print}\ ,\ \bibinfo {eid} {arXiv:2101.07045}} (\bibinfo {year}
  {2021})},\ \Eprint {http://arxiv.org/abs/2101.07045} {arXiv:2101.07045
  [cond-mat.str-el]} \BibitemShut {NoStop}%
\bibitem [{\citenamefont {{Xu}}\ \emph {et~al.}(2022)\citenamefont {{Xu}},
  \citenamefont {{Shi}}, \citenamefont {{Vitali}}, \citenamefont {{Qin}},\ and\
  \citenamefont {{Zhang}}}]{ZhangStripes}%
  \BibitemOpen
  \bibfield  {author} {\bibinfo {author} {\bibfnamefont {H.}~\bibnamefont
  {{Xu}}}, \bibinfo {author} {\bibfnamefont {H.}~\bibnamefont {{Shi}}},
  \bibinfo {author} {\bibfnamefont {E.}~\bibnamefont {{Vitali}}}, \bibinfo
  {author} {\bibfnamefont {M.}~\bibnamefont {{Qin}}}, \ and\ \bibinfo {author}
  {\bibfnamefont {S.}~\bibnamefont {{Zhang}}},\ }\href {\doibase
  10.1103/PhysRevResearch.4.013239} {\bibfield  {journal} {\bibinfo  {journal}
  {Physical Review Research}\ }\textbf {\bibinfo {volume} {4}},\ \bibinfo {eid}
  {013239} (\bibinfo {year} {2022})},\ \Eprint
  {http://arxiv.org/abs/2112.02187} {arXiv:2112.02187 [cond-mat.str-el]}
  \BibitemShut {NoStop}%
\bibitem [{\citenamefont {Xiao}\ \emph {et~al.}(2023)\citenamefont {Xiao},
  \citenamefont {He}, \citenamefont {Georges},\ and\ \citenamefont
  {Zhang}}]{Xiao2023}%
  \BibitemOpen
  \bibfield  {author} {\bibinfo {author} {\bibfnamefont {B.}~\bibnamefont
  {Xiao}}, \bibinfo {author} {\bibfnamefont {Y.-Y.}\ \bibnamefont {He}},
  \bibinfo {author} {\bibfnamefont {A.}~\bibnamefont {Georges}}, \ and\
  \bibinfo {author} {\bibfnamefont {S.}~\bibnamefont {Zhang}},\ }\href
  {\doibase 10.1103/PhysRevX.13.011007} {\bibfield  {journal} {\bibinfo
  {journal} {Phys. Rev. X}\ }\textbf {\bibinfo {volume} {13}},\ \bibinfo
  {pages} {011007} (\bibinfo {year} {2023})}\BibitemShut {NoStop}%
\end{thebibliography}
%

\end{document}